\documentclass[twocolumn]{aastex7}

\newcommand{\Msun}{M$_{\odot}$}
\newcommand{\Mbh}{$M_{\rm BH}$}
\newcommand{\Mnsc}{$M_{\rm NSC}$}

\newcommand{\Lsun}{L$_\odot$}

\newcommand{\ml}{\emph{M/L}}

\newcommand{\hst}{HST}
\newcommand{\jwst}{JWST}
\newcommand{\kms}{km~s$^{-1}$}

\newcommand{\go}{G$_{\rm outer}$}

\newcommand{\jamcyl}{\textsc{JAM}$_{\rm cyl}$}
\newcommand{\jamsph}{\textsc{JAM}$_{\rm sph}$}
\newcommand{\vrms}{$V_{\rm rms}$}
\newcommand{\jam}{\textsc{JAM}}
\newcommand{\mge}{\textsc{MGE}}
\newcommand{\ppxf}{\textsc{pPXF}}
\newcommand{\adamet}{\textsc{AdaMet}}
\newcommand{\osiris}{\textsc{OSIRIS}}
\newcommand{\powerbin}{\textsc{PowerBin}}
\newcommand{\mlhst}{$M/L_{\rm F547M}$}

\newcommand{\Vrms}{$\sqrt{V^2+\sigma^2}$}

\hypersetup{colorlinks, linkcolor=blue, citecolor=blue, urlcolor=blue}

\usepackage{multirow}
\usepackage{footnote}
\usepackage{newtxtext,newtxmath}
\usepackage{graphicx}	
\usepackage{amsmath}	
\usepackage{lineno}
\usepackage{caption}
\usepackage{xcolor}
\linenumbers

\graphicspath{{./}{figures/}}
\received{June 17, 2026}
\revised{October 30, 2026}
\accepted{November 10, 2026}
\submitjournal{ApJ}

\shorttitle{SMBH Measurements in NGC~5055 with Keck/OSIRIS}
\shortauthors{Tien H.\ T.\ Ho et al.}
\begin{document}

\title{The Black Hole Mass of the Sunflower Galaxy NGC 5055: Direct Measurement via Stellar Kinematics with Keck/OSIRIS Laser Guide Star Adaptive Optics}

\correspondingauthor{Dieu D.\ Nguyen}\email{dieun@umich.edu}
\author[0009-0005-8845-9725]{Tien H.\ T.\ Ho}
\email{htien2808@gmail.com}
\affiliation{Faculty of Physics—Engineering Physics, University of Science, VNU-HCM, Ho Chi Minh City, Vietnam}
\affiliation{Viet Nam National University Ho Chi Minh City, Vietnam}
\author[0000-0002-5678-1008]{Dieu D.\ Nguyen}
\email{dieun@umich.edu}
\affiliation{Department of Astronomy, University of Michigan, 1085 South University Avenue, Ann Arbor, MI 48109, USA}
\author[0000-0002-6496-9414]{Alister W.\ Graham}
\email{agraham@swin.edu.au}
\affiliation{Centre for Astrophysics and Supercomputing, Swinburne University of Technology, Hawthorn, VIC 3122, Australia}
\author[0000-0002-2126-3905]{Mark Durr\'e}
\email{mdurre@swin.edu.au}
\affiliation{Centre for Astrophysics and Supercomputing, Swinburne University of Technology, Hawthorn, VIC 3122, Australia}
\author[0009-0006-5852-4538]{Hai N.\ Ngo}
\affiliation{Department of Astronomy, University of Michigan, 1085 South University Avenue, Ann Arbor, MI 48109, USA}
\email{hainn@umich.edu}
\author[0009-0004-3689-8577]{Tinh Q.\ T.\ Le}
\email{lethongquoctinh01@gmail.com}
\affiliation{Department of Physics, International University, VNU-HCM, Ho Chi Minh City, Vietnam}
\affiliation{Viet Nam National University Ho Chi Minh City, Vietnam}
\author[0000-0001-5802-6041]{Elena Gallo}
\email{egallo@umich.edu}
\affiliation{Department of Astronomy, University of Michigan, 1085 South University Avenue, Ann Arbor, MI 48109, USA}
\author[0000-0002-7924-3253]{Igor V.\ Chilingarian}
\email{igor.chilingarian@cfa.harvard.edu}
\affiliation{Center for Astrophysics – Harvard and Smithsonian, 60 Garden St. Cambridge, MA, 02138 USA}
\affiliation{Observatoire de Paris, LUX, CNRS, PSL University, 75014, Paris, France}
\author[0009-0009-0015-1208]{Tuan N.\ Le}
\email{tuan.le.nutshell@gmail.com}
\affiliation{Faculty of Physics—Engineering Physics, University of Science, VNU-HCM, Ho Chi Minh City, Vietnam}
\affiliation{Viet Nam National University Ho Chi Minh City, Vietnam}
\author[0000-0002-4622-796X]{Roberto Soria}
\email{baobi123987@gmail.com}
\affiliation{INAF-Osservatorio Astrofisico di Torino, Strada Osservatorio 20, I-10025 Pino Torinese, Italy}
\affiliation{Sydney Institute for Astronomy, School of Physics A28, The University of Sydney, Sydney, NSW 2006, Australia}
\author[0000-0002-4436-6923]{Fan Zou}
\email{fanzou@umich.edu}
\affiliation{Department of Astronomy, University of Michigan, 1085 South University Avenue, Ann Arbor, MI 48109, USA}
\author[0009-0008-6050-5736]{Huy G.\ Tong}
\email{tonggiahuy191203@gmail.com}
\affiliation{Institute of Astronomy, National Tsing Hua University, Kuang Fu Rd. Sec. 2, Hsinchu, 30013, Taiwan}
\author[0009-0009-4259-1535]{Nghi K.\ N.\ Le}
\email{lengockhanhnghi@gmail.com}
\affiliation{Faculty of Physics—Engineering Physics, University of Science, VNU-HCM, Ho Chi Minh City, Vietnam}
\affiliation{Viet Nam National University Ho Chi Minh City, Vietnam}
\author[0000-0001-9649-2449]{Que T.\ Le}
\email{ltque@hcmiu.edu.vn}
\affiliation{Department of Physics, International University, VNU-HCM, Ho Chi Minh City, Vietnam}
\affiliation{Viet Nam National University Ho Chi Minh City, Vietnam}

\begin{abstract}
We present the first stellar-dynamical supermassive black hole mass measurement in the nearby spiral galaxy NGC~5055 (M63) using near-infrared integral-field spectroscopy with Keck/\osiris\ assisted by the laser guide star adaptive optics system. The observations deliver two-dimensional stellar kinematics extracted from the CO bandhead absorption features via \ppxf\ at an angular resolution of $0\farcs045$. The stellar velocity dispersion rises steeply from $\sim$90~\kms\ at $r \sim 1\arcsec$ to a central peak of $\sim$135~\kms, kinematically tracing the gravitational influence of a central mass concentration. Modeling these kinematics with an ensemble of four \jam\ models, spanning cylindrical and spherical velocity-ellipsoid alignments and two independent \mge\ decompositions of archival \hst\ imaging, yields a black hole mass of $M_{\rm BH}=4.56^{+1.73}_{-1.82}\times10^6$~\Msun. The stellar orbits are consistent with near-isotropy ($\beta \approx 0$), typical of rotating late-type galaxies. NGC~5055 emerges as a remarkable local analogue to the Milky Way in terms of its black hole, nuclear star cluster, and total stellar mass. The surface brightness profile of NGC~5055 shows no evidence for a classical bulge; if confirmed bulgeless, NGC~5055 would join the rare class of spirals with a direct dynamical \Mbh\ determination in the absence of a bulge, a regime currently populated by only a handful of galaxies including NGC~2748 and NGC~4395. This measurement provides a new anchor for black hole scaling relations in late-type galaxies at the low-mass end of the black hole mass spectrum.
\end{abstract}



\section{Introduction}\label{sec:intro}

While morphology-dependent coevolutionary scaling relations between supermassive black holes (SMBHs) and their host galaxies have emerged in recent years, this fundamental paradigm would benefit from more data at the low-mass end. In the regime of late-type and dwarf galaxies with \Mbh~$\lesssim 10^{6\text{--}7}$~\Msun, the precise mechanisms driving BH-galaxy co-evolution remain a subject of active and vigorous research. The tight correlations discovered between black hole (BH) mass and host-galaxy properties, such as the stellar velocity dispersion within the inner region ($\sigma_0$) or the effective half-light radius ($\sigma$), and the bulge luminosity ($L$), stellar mass ($M_\star$), dynamical mass, and radial concentration of stars \citep[e.g.,][]{Ferrarese00, Gebhardt00, Graham2001, McLure2002, Marconi03, Graham2012}, suggest one, or multiple, fundamental co-evolutionary links between SMBHs and their hosts over cosmic time.

Active galactic nuclei (AGN) feedback models propose that SMBHs regulate host-galaxy star formation via momentum- or energy-driven winds, providing a physical basis for part of these scaling relations \citep{Haehnelt1998, Silk98, King2003, DiMatteo2005}. However, the precise mechanisms driving this co-evolution remain a subject of active debate and may involve both energy and momentum conserving winds at different stages \citep{Silk2024}. At high masses, \citet{Graham2023b} explains how (cold gas)-poor mergers drive a dramatic departure, as dictated by the Virial theorem, from these relations predicted by AGN feedback.  At low masses, there is reason to believe that galaxies again deviate from the AGN-dictated scaling relations.  Indeed, given that dwarf early-type galaxies (dETGs) follow an $L_{\rm gal} \propto \sigma^{2.5}$ relation \citep[e.g.,][]{Davies1983, Toloba2014}, then if they follow a near-linear \Mbh\--$L$ relation \citep[e.g.,][]{Magorrian98}, they must follow an \Mbh\--$\sigma^{2.5}$ relation, placing them above-and-left of the extrapolated AGN-expectation for \Mbh\--$\sigma$ relation at low-masses. At the same time, some spiral galaxies have been observed to reside to the lower-right of this relation \citep{Graham2008, Hu2008}. Expanding the sample of direct, dynamical \Mbh\ measurements at low masses is therefore essential for discriminating between competing theoretical models and for calibrating secondary mass indicators \citep[see][for reviews]{Ferrarese2005, Kormendy13, Graham2016}.

Late-type galaxies (LTGs) are particularly valuable in this context. They span a wide range of morphologies, structural complexities, and star-formation histories, and they preferentially occupy the low-mass end of the \Mbh\ distribution, precisely the regime where competing co-evolution models make divergent predictions. Moreover, LTGs frequently host composite central structures shaped by both merger-driven and secular processes, making them sensitive probes of how \Mbh--host relations depend on assembly history. Yet among the $\sim$200 direct \Mbh\ measurements published to date \citep[e.g.,][]{Graham2008, Kormendy13, Saglia16, Bosch2016, Sahu19b, Sahu19a}, only several dozen reside in LTGs, introducing a significant morphological bias into the current census.

Obtaining dynamical \Mbh\ in LTGs is observationally challenging for two primary reasons. First, the \Mbh\ predicted from scaling relations is relatively small ($\lesssim$$10^6$--$10^7$~\Msun\ for $\sigma \lesssim 100$~\kms), such that the gravitational sphere of influence \citep[SOI\footnote{The region surrounding the BH within which its gravitational potential dominates the stellar dynamics, defined by $R_{\rm SOI} = GM_{\rm BH}/\sigma^2$, where the enclosed stellar mass equals approximately \Mbh.};][]{Merritt2001, Merritt2004} typically subtends only a few tens of milliarcseconds (mas) at distances of $\sim$5--15~Mpc \citep{Ferrarese2005}. Second, the prominent disk and dust features in LTGs complicate optical surface-brightness photometry and can bias dynamical mass models. These obstacles are increasingly overcome through (i)~near-infrared (NIR) integral-field spectroscopy (IFS) with laser-guide-star adaptive optics (LGS-AO), which provides $\lesssim$$0\farcs1$ resolution while penetrating dust extinction; and (ii)~Jeans Anisotropic Modeling \citep[\jam\footnote{v7.2.4: \url{https://pypi.org/project/jampy/}};][]{Cappellari08, Cappellari2020} paired with Multi-Gaussian Expansion \citep[\mge\footnote{v6.0.4: \url{https://pypi.org/project/mgefit/}};][]{Emsellem94, Cappellari2002} photometric modeling. \jam\ efficiently models two-dimensional (2D) stellar kinematic maps using axisymmetric Jeans equations that account for a central dark object, orbital anisotropy, and the bulk of the galaxy's mass distribution. Such methods have yielded robust \Mbh\ measurements in ETGs \citep[e.g.,][]{Nguyen17, Nguyen18, Nguyen19, Nguyen23, Nguyen_2026_highz, Thater19, Thater23} and, more recently, in a growing number of LTGs \citep[e.g.,][]{Cappellari2009, Drehmer2015, Nguyen_2025_4736, Nguyen_2025_NSC, Nguyen_2026_4258, Nguyen_2026_M81, LeTinh2026}.

\subsection{The Target NGC~5055}\label{sec:targets}

NGC~5055 (M63, aka the ``Sunflower Galaxy'') is a nearby Sbc spiral galaxy in the small M51 Group of galaxies.  NGC~5055 is $\sim$0.3~mag fainter than M51 (aka the ``Whirlpool Galaxy"), with the next most dominant galaxy (NGC~5023, if ignoring M51B) some 3.6 mag fainter.  With two dominant spiral galaxies, it therefore has some similarity to the Local Group, which is dominated by M31$+$M32 and the Milky Way. 

NGC~5055 has a flocculent appearance at optical wavelengths \citep{Elmegreen1987} but displays a weak grand-design two-arm spiral ($60\arcsec < R_{\rm major} < 190\arcsec$) in the $K^{\prime}$-band \citep{Thornley1997}. Signs of minor disturbances are evident in optical images from, for example, The Virtual Telescope Project\footnote{\url{https://www.virtualtelescope.eu/}} \citep{Masi2014} and extensive tidal tails and asymmetries in the galaxy emerge in deep images \citep{Martinez-Delgado2010, Morales2018}. NGC~5055 has a precisely determined Tip-of-the-Red-Giant-Branch (TRGB) distance of $D = 8.87 \pm 0.31$~Mpc \citep{McQuinn_2017}, corresponding to a physical scale of $43$~pc~arcsec$^{-1}$ under a flat $\Lambda$CDM cosmology with $H_0 = 70$~\kms~Mpc$^{-1}$, $\Omega_{\Lambda} = 0.7$, and $\Omega_{\rm m} = 0.3$. At this distance, the logarithm of the black hole mass is predicted to be 6.81$\pm$0.63~dex (\Mbh~$= 6.53\times10^6$~\Msun), when using the \Mbh--$\sigma$ relation for LTGs \citep[][their equation~2]{Sahu19b} and $\sigma = 117(\pm6)$~\kms\ \citep{Ho_2009}. This yields $R_{\rm SOI} \simeq 0\farcs069$ ($=2.96$~pc), a factor of 1.5 above the resolution limit of our Keck/\osiris\ LGS-AO data, which has a full width at half maximum of the point spread function of ${\rm FWHM_{PSF}^{OSIRIS}} = 0\farcs045$ (Table~\ref{tab:mge_psf}). 

Previous attempts to constrain the central mass in NGC~5055 lacked the spatial resolution to isolate the BH's gravitational influence from the surrounding stellar potential. \citet{Perez_1995} derived a total dynamical enclosed mass of $M_{\rm encl} \sim 1.5\times10^8$~\Msun\ within the central $\pm5\arcsec$ ($\approx$215~pc) from ground-based long-slit spectroscopy, also revealing a linear rotation curve and a $\sim$80~\kms\ outflow in this region. From CFHT Fabry-Perot H$\alpha$ observations, \citet{Blais-Ouellette1999} identified anomalous kinematics consistent with a compact rotating disk of radius $\sim$8\farcs5 ($\approx$365~pc) and inferred a central dark mass of $\sim$$10^{7.2}$--$10^{7.5}$~\Msun\ via a simple Keplerian estimate. While these two estimates are not contradictory: the former is the total enclosed mass dominated by stars and gas, the latter a point-like dark object, neither resolves the BH SOI, which is expected to extend only to $R_{\rm SOI} \lesssim 2$~pc (Section~\ref{sec:results_discussion}). No direct stellar-dynamical \Mbh\ measurement has since been reported for NGC~5055.

The nuclear region is otherwise well characterized. The nucleus is classified as a low-ionization nuclear emission-line region galaxy \citep[LINER;][]{Heckman_1980, Ho_2008} and exhibits signatures of low-luminosity AGN activity, including compact X-ray emission and UV stellar-wind features \citep{Maoz1995, Maoz_1998}. IFS with the 6~m SAO/RAS telescope by \citet{Afanasiev_Silchenko_2002} identified a kinematically distinct circumnuclear stellar disk and a lopsided, Mg-enhanced substructure displaced $\sim$100~pc ($\sim$2.3$\arcsec$) south-west of the photometric center, with mean stellar ages of 3--4~Gyr, significantly younger than the surrounding populations (8--10~Gyr), suggesting episodic star formation driven by secular gas inflow.  On larger scales, \ion{H}{1} interferometry reveals a symmetric, warped gaseous disk extending to $\sim$40~kpc \citep{Battaglia_2006}, while the {\it Spitzer} Survey of Stellar Structure in Galaxies (S$^4$G) photometry provides a well-constrained stellar mass-to-light ratio for dynamical modeling \citep[$M/L_{3.6\,\mu\rm m} \approx 0.55$~\Msun/\Lsun;][]{Jimenez_Campos_2017}. NGC~5055 thus represents a nearby, multiwavelength-rich target lacking a stellar-dynamical \Mbh\ determination. 

Crucially, with a stellar mass of $\log(M_{\rm \star,gal}/M_\odot) \approx 10.81$ \citep{Davis2018}, NGC~5055 provides an opportunity to study a local, late-type structural analogue to the Milky Way ($\log (M_{\rm \star,gal}/{\rm M}_\odot)=10.8$~dex), allowing us to better place our own galaxy's evolutionary history into a broader cosmological context.

\subsection{This Work}\label{sec:thiswork}

We present the first stellar-dynamical measurement of \Mbh\ in NGC~5055 using NIR IFS observations obtained with \osiris\ behind the LGS-AO system on the 10-m Keck~I telescope. With a PSF FWHM of 0\farcs045, our data achieve a physical resolution of $\sim$1.9~pc, providing the highest-resolution view of this galaxy's central region to date. Two-dimensional line-of-sight (LOS) velocity distribution (LOSVD) of the stellar kinematics is extracted from the CO absorption bandheads ($\sim$2.3~\micron) using the Penalized PiXel-Fitting (\ppxf\footnote{v8.2.1: \url{https://pypi.org/project/ppxf/}}) method \citep{Cappellari2004, Cappellari2017}. These kinematics are modeled using \jam\ \citep{Cappellari08, Cappellari2020} in conjunction with \mge\ decompositions of archival {\it Hubble Space Telescope} (\hst) and ground-based $V$-band imaging that characterize the surface brightness profile from pc to kpc scales \citep{Fisher2008, Pechetti2020}.  

This paper is organized as follows. Section~\ref{sec:osiris_observations} describes the Keck/\osiris\ observations, data reduction, and PSF characterization. Sections~\ref{sec:losvd} and \ref{sec:photomodel} present the stellar kinematic extraction and photometric modeling of NGC~5055, respectively. Section~\ref{sec:dynamical-models} describes the \jam\ dynamical modeling. Results and discussion are presented in Section~\ref{sec:results_discussion}, and we summarize our conclusions in Section~\ref{sec:conclusions}.

\section{\osiris\ Observations and Data Reduction} \label{sec:osiris_observations}

\subsection{IFS Observations and Data Reduction}\label{sec:obs}

We observed the nucleus of NGC~5055 with the OH Suppressing InfraRed Imaging Spectrograph  \citep[\osiris;][]{Larkin2006} with the Keck I LGS-AO system \citep{vanDam2006,Wizinowich2006}. \osiris{} is a lenslet array IFU spectrograph with a $2048 \times 2048$ Hawaii-2 detector and spectral resolution in the \textit{K} band of R $\sim$ 3600. The observations were conducted on 15 Jan. 2025, and consisted of two 900-second object frames, with an intervening 900-second sky frame (i.e., 30 min on object), in the standard object-sky-object sequence, at an airmass of 1.08. The plate scale was 0\farcs035, which at the distance of NGC~5055 (8.87 Mpc) equates to 1.5 pc pixel$^{-1}$. We observed with the $Kn5$ filter (2.292-2.408 $\mu$m), which is ideally suited for CO bandhead observations. 

Each object frame was reduced using the \osiris{} data reduction pipeline (DRP) version 6.0, using the sky frames for background subtraction. The DRP uses ``Rectification Matrices" which incorporate all the flat fielding, bias and dark frames in one, along with mapping each pixel to a lenslet/spectral element. These matrices are retrieved from the Keck repository; for our data, these were dated Feb. 2025. The resulting data cubes have fields of view of 1\farcs785 $\times$ 2\farcs31 (51 $\times$ 66 pixels). 

The data cubes were centered and combined, then trimmed for display purposes, resulting in images with about 2100 spectra. The resulting data cube values were cleaned of spikes and geometric artifacts; these are mainly a product of the DRP. Telluric correction was performed with observation of the A0V star, HIP~61471. The stellar spectrum is interpolated over the hydrogen absorption features and a representative blackbody is assumed. This is divided through the data cube spectra to remove the telluric absorption features. Flux calibration is performed by integrating the stellar spectrum over a clean region (away from telluric and stellar features), then matched to a reference Vega spectrum.

\subsection{\osiris\ Point Spread Functions}\label{sec:osirsK_psf}

Since the empirical \osiris\ PSF depends on the AO correction quality and atmospheric conditions at the time of observation, we derived it directly from a standard star observed contemporaneously with the NGC~5055 science frames. We observed HIP~61471 with the AO loops closed immediately following the science exposures in a short exposure time to avoid saturation, using the same plate scale ($35\,\rm{mas}\,\rm{pixel}^{-1}$) and filter (\textit{Kn5}). The PSF image was constructed from the normalized, median-combined data cube collapsed over $2.295$--$2.305\,\micron$, a wavelength range within the \textit{Kn5} bandpass and matching the spectral region used for stellar kinematic extraction. This empirical PSF thus captures the instantaneous AO performance during the NGC~5055 observations.

The PSF image was parameterized as a Multi-Gaussian Expansion \citep[\mge;][]{Emsellem94} using the \texttt{mge\_fit\_sectors} routine from the \textsc{MgeFit}\footnote{v5.0.14: \url{https://pypi.org/project/mgefit/}} package \citep{Cappellari2002}. The resulting \mge\ model is listed in Table~\ref{tab:mge_psf} and is convolved with all \jam\ model predictions to account for the \osiris\ PSF at $2.3\,\micron$ (Section~\ref{sec:dynamical-models}).  The azimuthally averaged radial profile of HIP~61471 and this corresponding MGE fit are shown in Figure~\ref{fig:star_psf_mge}, with the PSF FWHM and SOI indicated for reference.

\section{Stellar Kinematics Measurements}\label{sec:losvd}

\subsection{Defining the Galaxy Center}\label{sec:center}

Due to the uncertainties in LOSVD extractions (Section~\ref{sec:kinemap}), we adopt the photometric center as the coordinate origin, assuming the central BH coincides with the surface-brightness peak in the \osiris\ white-light image. To determine the weighted barycenter of the nucleus, we employed a two-step refinement: first using the 200 brightest pixels within a $0\farcs35 \times 0\farcs7$ aperture of the \osiris\ white-light image, then the 60 brightest pixels of its $0\farcs21 \times 0\farcs35$ region. The estimates agree within a quarter of a pixel ($\Delta X < 0\farcs018$, $\Delta Y < 0\farcs017$), a difference negligible at the spatial resolution of the Keck/\osiris\ IFU data.  Consequently, we adopt the barycenter from the larger aperture (R.A., Decl. = $13^{\text{h}}15^{\text{m}}49\text{.}^{\text{s}}155$, $+42^{\circ}01'46\farcs86$; consistent with the HST/WFPC2 F547M photometric center from \citealt{Pechetti2020}) determination for its superior statistical robustness and used it as the origin in all subsequent models and figures.

\begin{figure*}
    \centering
    \includegraphics[width=0.8\textwidth]{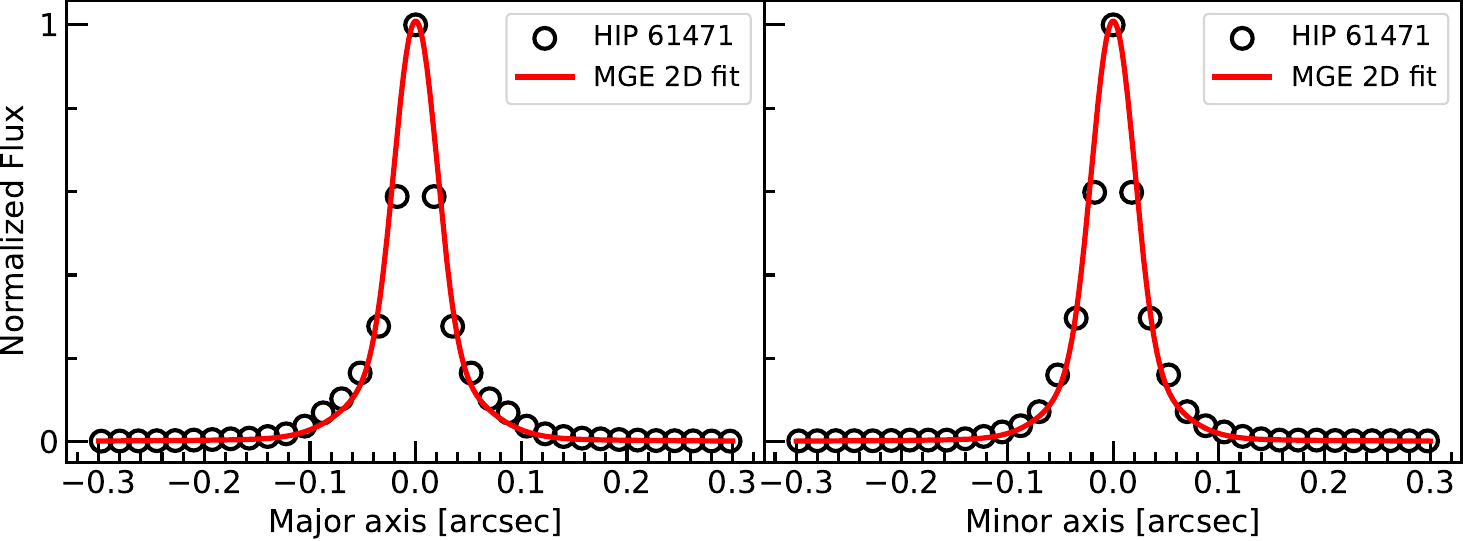}
    \caption{Keck/OSIRIS $Kn5$ PSF constructed from the star HIP~61471, which is extracted along the major (left) and minor (right) axes. Black open circles show the data and the red line is its best-fit 2D MGE model (Table~\ref{tab:mge_psf}). At $r \sim 0\farcs1$, the MGE fit reproduces the PSF profile to within 8\% uncertainty.} 
    \label{fig:star_psf_mge}
\end{figure*}

\begin{table}
\centering
\caption{Keck/\osiris\ PSF MGE models (HIP~61471) \label{tab:mge_psf}}
\vspace{-3mm}
\begin{tabular}{cccc}
\hline\hline
$j$&(Light fraction)$_j$&$\sigma_j~({\rm arcsec})$&FWHM$^{\rm OSIRIS}_{\rm PSF}~({\rm arcsec})$ \\
(1) & (2) & (3) & (4) \\
\hline
1 & 0.805 & 0.017 & \multirow{3}{*}{0.045} \\
2 & 0.187 & 0.040 & \\
3 & 0.008 & 0.150 & \\
\hline
\end{tabular}
\noindent\parbox{\linewidth}{\textbf{Notes.} (1) Gaussian component index; (2) fractional contribution to the total light; (3) Gaussian dispersion along the major axis; (4) intensity-weighted mean FWHM of the PSF.}
\end{table}

\subsection{\ppxf\ Setup \& The Optimal Stellar Template}\label{sec:optimal_template}

Stellar kinematics for NGC~5055 were extracted from the Keck/\osiris\ $Kn5$-band data cube using the \ppxf\ method \citep{Cappellari23}, primarily targeting the CO bandhead absorption features \citep[e.g.,][]{Cappellari2009, Nguyen14, Ngo_2025_NSC, Ngo2025c}. Following the results of \citet{Nguyen_2026_4258, Nguyen_2026_M81}, who found that LOSVDs derived from the empirical X-shooter Spectral Library \citep[XSL\footnote{DR3 with 830 spectra from 683 stars:\url{http://xsl.u-strasbg.fr}};][]{Verro2022} agree within 3\% with those from the higher-resolution PHOENIX synthetic library \citep{Husser2013}, we adopt XSL as our primary template source. The XSL provides broad spectral coverage (3000--25,000~\AA) at $R\sim10,000$, encompassing a diverse range of O--M and evolved stars.

\begin{figure*}
 \centering
   	\includegraphics[width=0.7\linewidth]{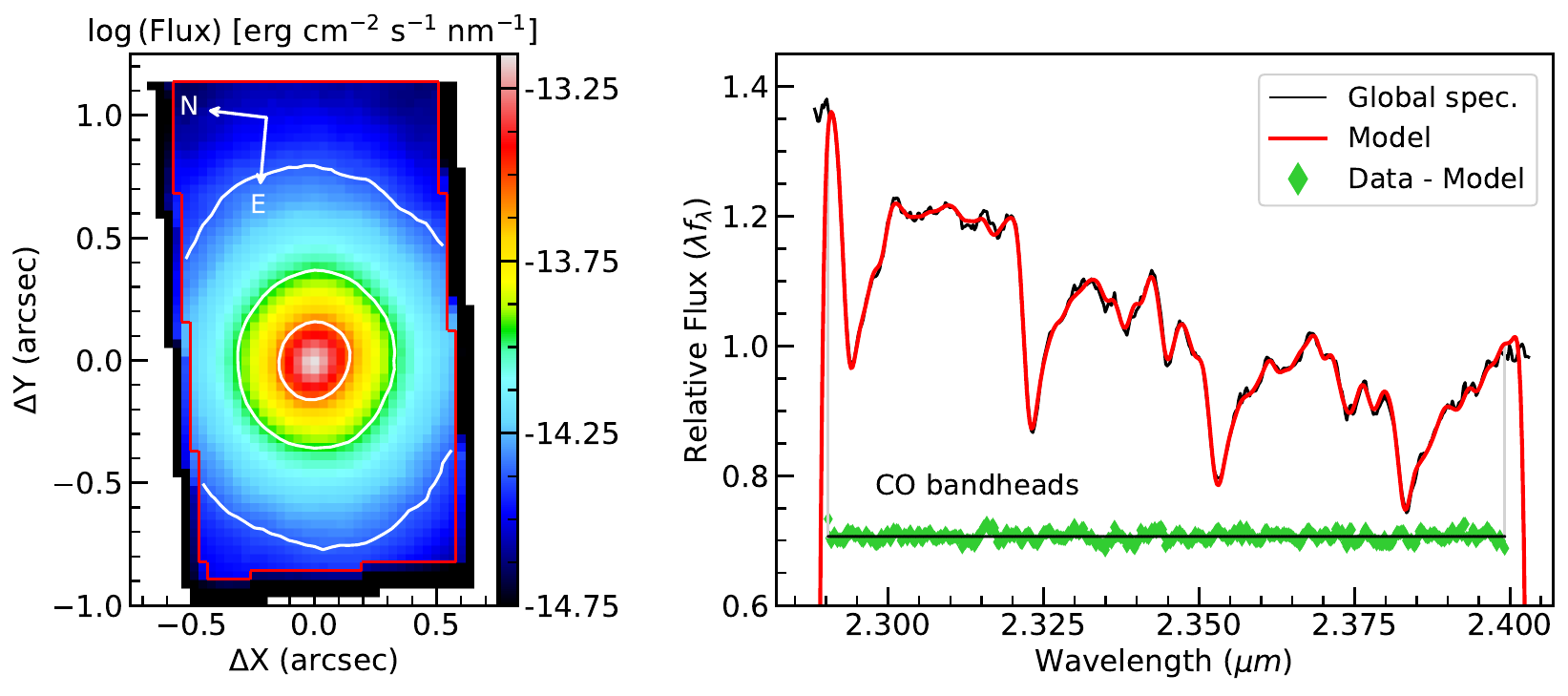} 
	\caption{{\it Left:} Logarithmic map of the integrated stellar continuum from the Keck/\osiris\ data cube, collapsed along the spectral axis. {\it Right:} Optimal stellar template fit to the integrated spectrum of NGC~5055 by co-adding all spaxels within the red aperture shown in the left panel. The observed spectrum (black) is overlaid with the best-fit \ppxf\ broadened template (red); residuals are shown as green dots. White contours trace the stellar continuum intensity in steps of 1 mag arcsec$^{-2}$.}
	\label{fig:Intensity_maps}
\end{figure*} 

To account for mixed stellar populations in the nucleus of NGC~5055 and mitigate kinematic biases from template mismatch \citep{Marconi2000, Silge2003}, we constructed an optimal XSL template from a large spatial region's integrated spectrum, namely the global spectrum.  Additive Legendre polynomials were excluded from all \ppxf\ fits (\texttt{degree} $= -1$), as they can modify the equivalent widths of the CO bandhead absorption features in the \textit{Kn5}-band and systematically bias the recovered LOSVDs. Multiplicative polynomials (\texttt{mdegree}) of orders 0--4 were instead explored to account for residual continuum shape, sky subtraction, and calibration systematics; the recovered kinematics converge and become insensitive to polynomial order at \texttt{mdegree} $\geq 2$ over the full fitting window ($2.29$--$2.40~\micron$). We therefore adopt \texttt{mdegree} $= 2$ as our fiducial configuration.

The procedure is as follows. First, we constructed a global spectrum by co-adding all spaxels within the red aperture shown in the left panel of Figure~\ref{fig:Intensity_maps} from the \osiris\ data cube, weighted by the continuum flux of each spaxel. The resulting spectrum reaches a signal-to-noise ratio of $\mathrm{S/N} \sim 316$ per spectral pixel and was logarithmically rebinned to a constant velocity scale of 32~\kms\ per pixel following Eq.~8 of \citet{Cappellari2017}. We broadened the XSL templates to match the instrumental resolution by convolving them with a Gaussian kernel with dispersion computed from Eq.~5 of \citet{Cappellari2017}.

\begin{table} 
\centering
\caption{Keck/\osiris\ Kinematic Data of the NGC 5055 Nucleus}
\vspace{-3mm}
\begin{tabular}{cccccc}
\hline \hline
$\Delta$R.A. &$\Delta$Decl. & $V$ & $\Delta V$ & $\sigma$ & $\Delta\sigma$ \\
(arcsec) & (arcsec) & (\kms) & (\kms) & (\kms) & (\kms) \\
\hline
 0.001  &$-$0.001 &$-$6.10  & 3.47 & 134.67 & 3.23 \\ 
 0.018  &$-$0.036 &  14.73  & 3.04 & 126.98 & 2.97 \\ 
 0.018  &   0.052 &$-$15.32 & 3.00 & 126.07 & 3.01 \\ 
$-$0.052&$-$0.018 &   0.96  & 2.96 & 126.06 & 2.93 \\ 
 0.059  &$-$0.013 &   6.14  & 3.13 & 121.43 & 3.21 \\ 
$-$0.043&   0.069 &$-$30.12 & 3.23 & 126.02 & 2.99 \\ 
$-$0.011&$-$0.083 &  23.98  & 2.53 & 121.75 & 2.73 \\ 
...     &    ...  &    ...  &  ... &  ...   & ... \\[1mm] 
\hline
\end{tabular}
\noindent\parbox{\linewidth}{(This table is available in its entirety in machine-readable form)}
\label{tab:kin}
\end{table}

\begin{figure*}[!th]
    \centering\includegraphics[width=0.7\textwidth]{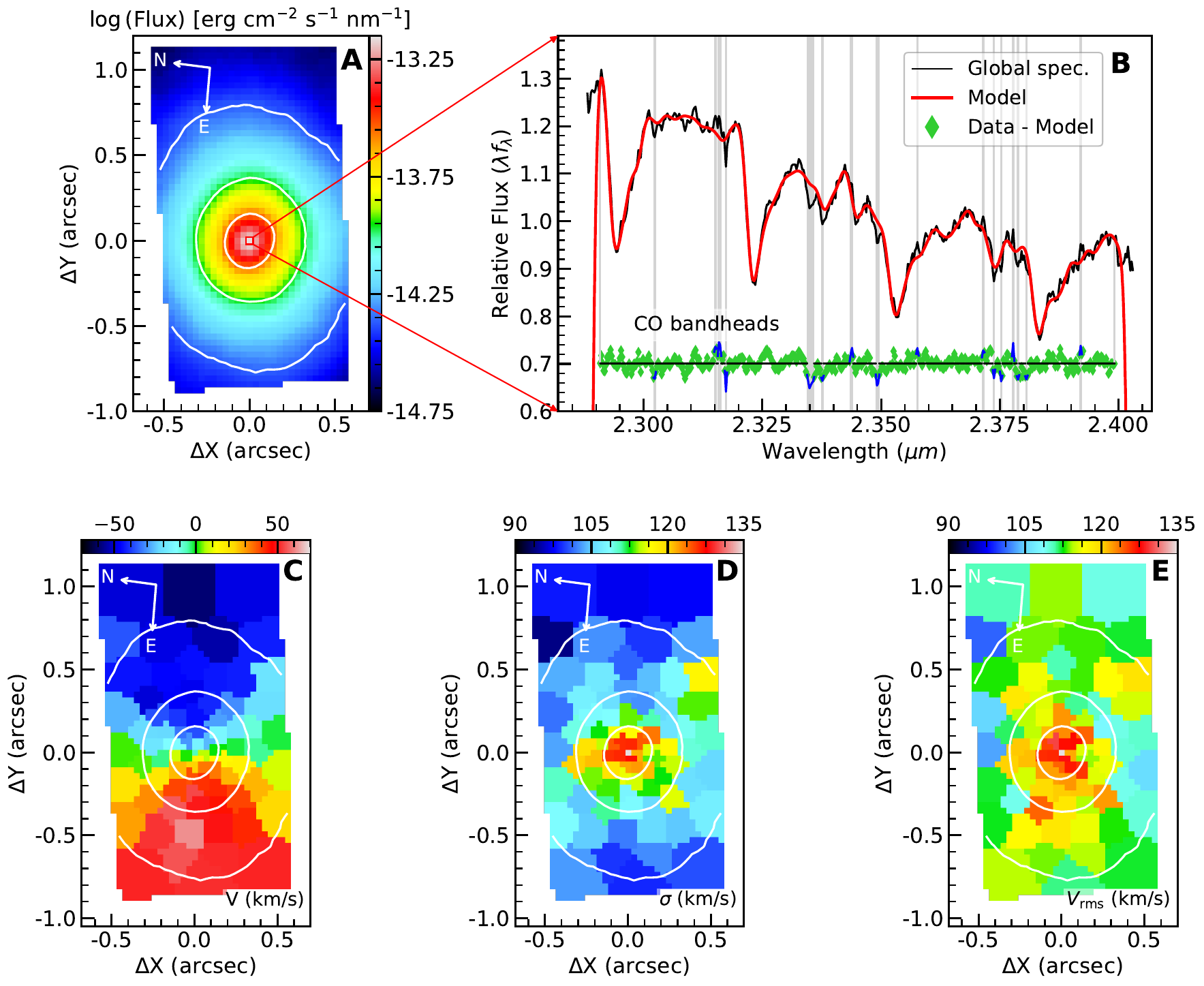}
    \caption{Stellar kinematics of NGC~5055 from Keck/\osiris. {\it Panel A:} Logarithmic integrated intensity map of the data cube. {\it Panel B:} Representative pPXF fit for a central-offset \powerbin\ bin (location marked in Panel A). The observed spectrum (black) shows CO bandhead absorptions near 2.3~\micron; the best-fit XSL template is overplotted in red. Residuals \texttt{data-model} are shown in green; vertical gray lines delimit the fitting window applied uniformly across all bins.  {\it Panels C--E:} 2D maps of the stellar line-of-sight velocity ($V$), velocity dispersion ($\sigma$), and root-mean-square velocity ($V_{\rm rms} =\sqrt{V^2+\sigma^2}$). White contours trace the stellar continuum intensity in steps of 1 mag arcsec$^{-2}$.}
    \label{fig:kin-map}
\end{figure*}

Second, we derived a fixed optimal XSL template by fitting the broadened templates to the global spectrum with \ppxf, adopting a Gaussian LOSVD (i.e., \texttt{moments} = 2). This yields the stellar kinematics, namely the rotational velocity $V$ (relative to the systemic velocity $V_{\rm sys}$) and the velocity dispersion $\sigma$, which are the primary quantities of interest.

The optimal template consists of 15 giant-star spectra that reproduce the observed features over 2.29--2.40~\micron, including the CO bandhead absorption (Figure~\ref{fig:Intensity_maps}, right) of the \osiris\ spectra. The fit is dominated by HD 161817 (XSL: X0639), contributing $\approx$38\% of the flux, with additional contributions from a giant K-type star (HD~38529B; $\approx$18\%) and an M-type star (BMB~106; $\approx$14\%). The best-fit integrated stellar kinematics yield a heliocentric velocity $V = 502 \pm 3$~\kms\ and velocity dispersion $\sigma = 108 \pm 3$~\kms, where uncertainties represent the $1\sigma$ scatter from 200 Monte Carlo realizations \citep{Cappellari2004}. Both values are in good agreement with independent measurements: $V$ is consistent with the \ion{H}{1} 21~cm systemic velocity of \citet{Springob2005}, and $\sigma$ agrees with the CaT and \ion{Mg}{1}b absorption-line dispersion measured with the Double Spectrograph at Palomar Observatory \citep{Barth2002}.

\begin{figure}
    \includegraphics[width=0.45\textwidth]{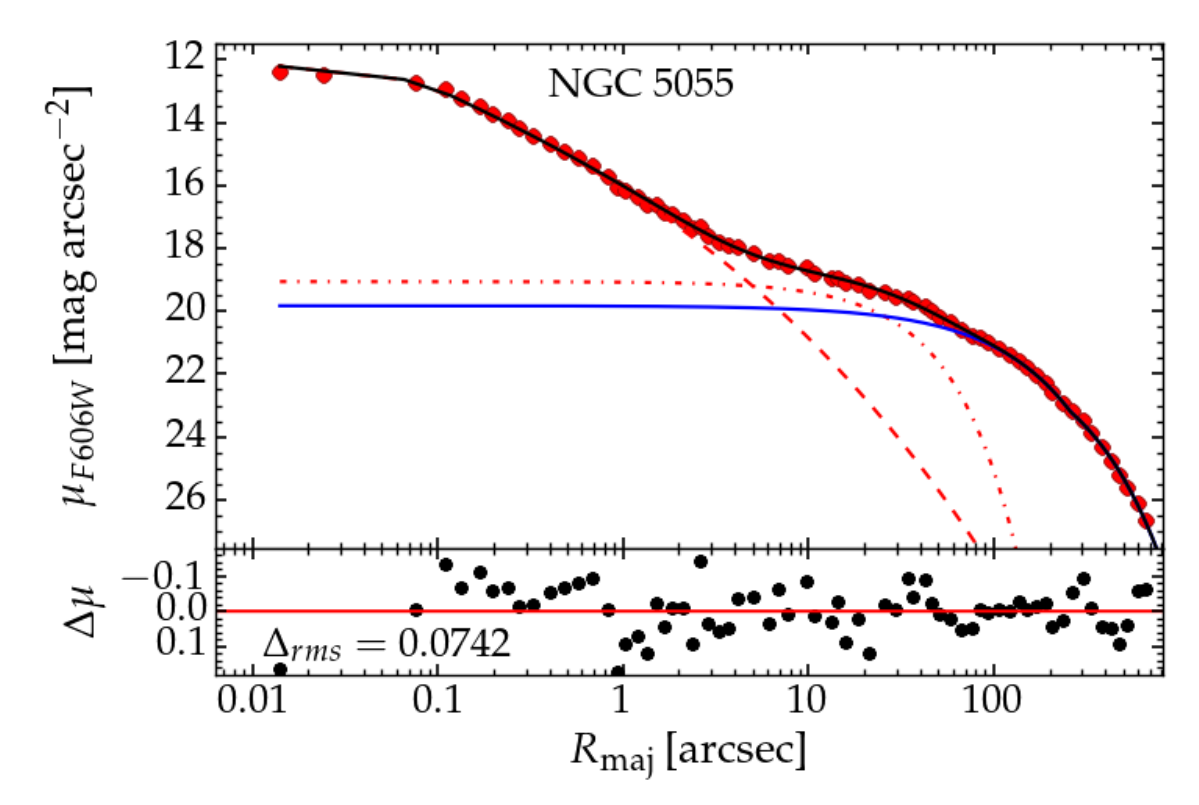}
    \caption{Decomposition of the surface brightness profile of NGC~5055 using the composite $V$-band major-axis profile from \citet[][their figure~16]{Fisher2008}, that combined HST/WFPC2 PC F606W imaging in the central region with wide-field ground-based photometry at large radii. \textit{Top:} The observed profile (filled circles) is plotted against $\log R_{\rm maj}$ and overlaid with a three-component model (solid black line), comprising an inner S\'ersic component (short-dashed red curve), a middle S\'ersic component (dot-dashed red curve), and an anti-truncated exponential disk (solid blue curve) that dominates at large radii. \textit{Bottom:} Residuals ($\Delta\mu$ = {\tt data $-$ model}) in mag~arcsec$^{-2}$.}
    \label{fig:ngc5055_f606w_SB}
\end{figure}

\begin{table*}[!t]
\centering
\caption{\hst/WFPC2 PC F547M and F606W (plus $V$-band ground-based photometry) \mge\ models}    
\begin{tabular}{cccc|ccc} 
\hline\hline
$j$ &$\Sigma_{\star,j}$ (\Lsun\ ${\rm pc^{-2}})$ &$\sigma_j$ (arcsec) &$q'_j=b_j/a_j$&$\Sigma_{\star,j}$ (\Lsun\ ${\rm pc^{-2}})$ &$\sigma_j$ (arcsec) &$q'_j=b_j/a_j$\\
(1) & (2) & (3) & (4)& (2) & (3) & (4)\\ 
\hline
&  \multicolumn{3}{c}{\citet{Pechetti2020}}&\multicolumn{3}{c}{Our refit \citet{Fisher2008}}\\
\hline
1  & 636972 & 0.002 & 0.87 & 275042 & 0.050   & 0.87\\ 
2  & 575440 & 0.007 & 0.87 & 103922 & 0.115   & 0.87\\ 
3  & 407380 & 0.021 & 0.87 & 43763  & 0.225   & 0.61\\ 
4  & 224400 & 0.054 & 0.87 & 10564  & 0.345   & 0.61\\ 
5  & 65700  & 0.072 & 0.87 & 15921  & 0.490   & 0.61\\ 
6  & 92700  & 0.127 & 0.87 & 8129   & 0.880   & 0.61\\ 
7  & 39100  & 0.138 & 0.87 & 3182   & 1.522   & 0.61\\ 
8  & 9430   & 0.196 & 0.87 & 1551   & 2.498   & 0.61\\ 
9  & 28600  & 0.273 & 0.61 & 760    & 4.522   & 0.61\\ 
10 & 23500  & 0.276 & 0.61 & 409    & 9.920   & 0.61\\ 
11 & 15300  & 0.488 & 0.61 & 513    & 27.596  & 0.61\\ 
12 & 5830   & 0.621 & 0.61 & 23     & 32.443  & 0.61\\ 
13 & 4120   & 0.821 & 0.61 & 102    & 69.874  & 0.61\\ 
14 & 4290   & 0.957 & 0.61 & 7      & 82.874  & 0.61\\ 
15 & 3320   & 1.713 & 0.61 & 69     & 111.568 & 0.61\\ 
16 & 964    & 3.844 & 0.61 & 28     & 169.309 & 0.61\\ 
17 & 1130   & 6.296 & 0.61 & 3      & 247.063 & 0.61\\[1mm]
\hline
\end{tabular}
\parbox[t]{\textwidth}{\small{\bf Notes:} (1) Gaussian component index; (2) Luminosity surface density; (3) Gaussian dispersion along the major axis; (4) Observed axial ratio.} 
\label{table:mge-table}
\end{table*}

\subsection{Two-Dimensional Stellar Kinematic Maps}\label{sec:kinemap}

Two-dimensional stellar LOSVDs across the nucleus of NGC~5055 were extracted as follows. We first applied the adaptive \powerbin\footnote{v1.1.11: \url{https://pypi.org/project/powerbin/}} algorithm \citep{Cappellari2025} to spatially bin spaxels to a target S/N of 162 per spectral pixel, yielding $N = 82$  bins. Spaxels beyond $r \approx 0\farcs048 \approx R_{\rm SOI}$ were grouped adaptively, while those interior to $R_{\rm SOI}$ were left unbinned to preserve the spatial resolution required for kinematic measurements within the SOI; individual unbinned spaxels there have S/N ranging from 162 to 165 per spectral pixel, increasing toward the nucleus. Each binned spectrum was resampled onto a logarithmic wavelength grid and fitted with \ppxf\ \citep{Cappellari2017} using the fixed optimal XSL template (Section~\ref{sec:optimal_template}) broadened to the instrumental resolution, with second-degree multiplicative Legendre polynomials to correct for residual spectral shape differences, yielding the stellar LOS velocity $V$ and dispersion $\sigma$.

Figure~\ref{fig:kin-map} presents the logarithmic integrated intensity map (panel A), with a marked spaxel corresponding to the spectrum and \ppxf\ fit shown in panel B. The remaining panels display the resulting maps of rotational velocity $V$ (panel C), velocity dispersion $\sigma$ (panel D), and root-mean-square velocity $V_{\rm rms} =$~\Vrms\ (panel E). The nucleus exhibits a rotation with $\vert V \vert \approx 70 \pm 4$~\kms\ at the edge of the \osiris\ FOV (after subtracting the systemic heliocentric velocity of $v_{\rm sys} = 500 \pm 2$~\kms).  This derived $v_{\rm sys}$ from the Keck/\osiris\ data cube is identical to the value reported in the NASA/IPAC Extragalactic Database (NED\footnote{\url{https://ned.ipac.caltech.edu}}).

We measure a global kinematic position angle of PA$_{\rm kin} = 83^\circ \pm 10^\circ$. Both $v_{\rm sys}$ and PA$_{\rm kin}$ were derived from the velocity map using the \textsc{pafit}\footnote{v2.0.8: \url{https://pypi.org/project/pafit/}} package \citep{Krajnovic2006}. The kinematic PA$_{\rm kin}$ is consistent within the uncertainties with the photometric one PA$_{\rm phot}\approx 99^\circ$ reported in the HyperLeda database\footnote{\url{https://leda.univ-lyon1.fr/}} \citep{Wevers1986}, as expected for a nearly axisymmetric disk galaxy. The $\sigma$ profile, which dominates over $V$, rises from $\sim$$108 \pm 3$~\kms\ at $r \gtrsim 0.068\arcsec$, to a central peak of $\sim$$135 \pm 3$~\kms\ within $r \lesssim 0\farcs1$. This central increase in both $\sigma$ and \vrms\ provides strong evidence for the presence of a central SMBH. We presented these stellar-kinematic measurements in Table~\ref{tab:kin}.

\section{Galaxy Photometric Model}\label{sec:photomodel}

\begin{figure*}[!th]
    \centering
    \includegraphics[height=0.8\textwidth]{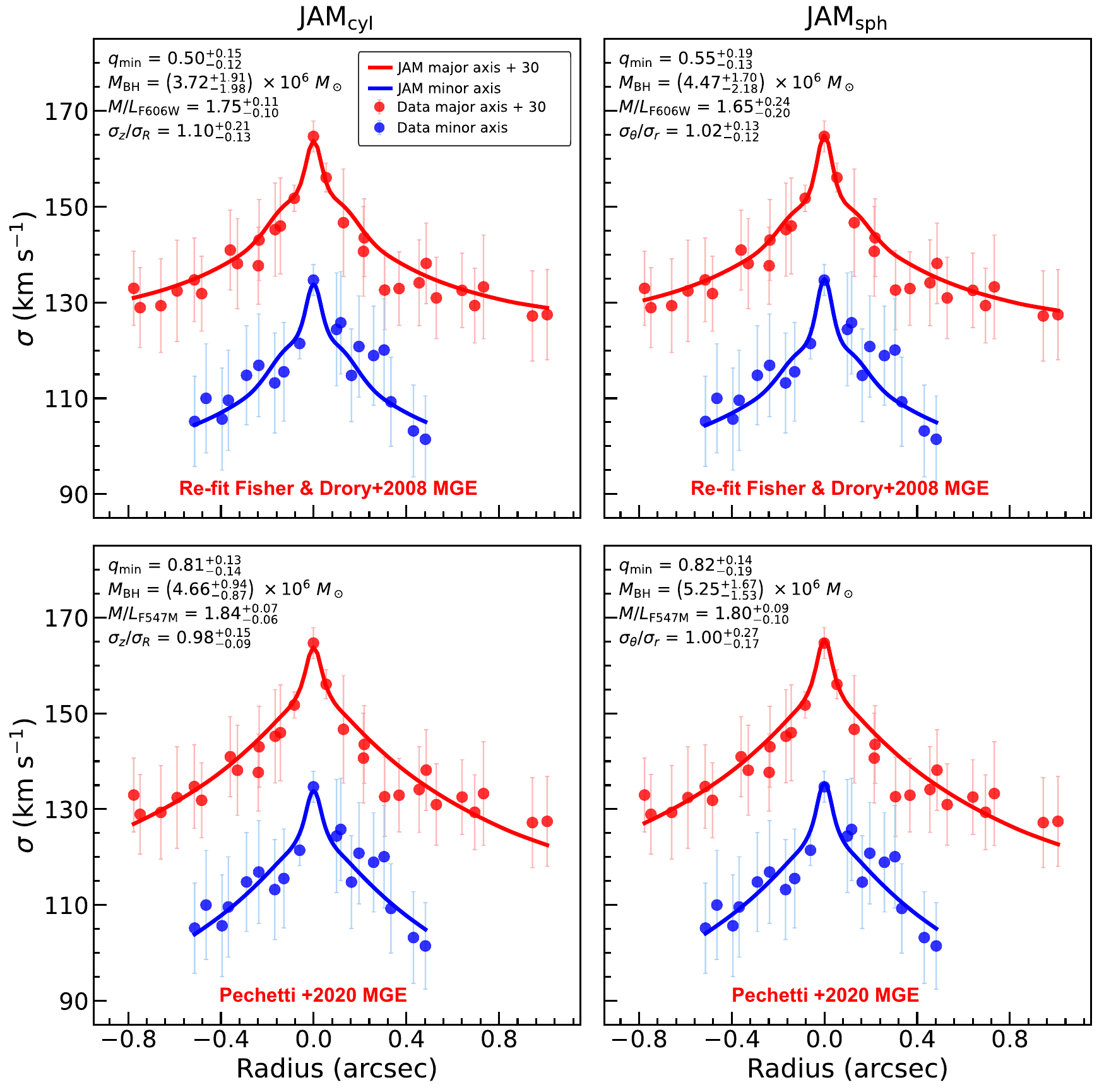}
    \caption{Stellar kinematic profiles of NGC~5055 extracted from the Keck/\osiris\ data cube (filled circles), overlaid with the best-fit \jamcyl\ (\textit{left}) and \jamsph\ (\textit{right}) model predictions (solid curves). Red and blue points correspond to the major- and minor-axis bins, respectively, with the major-axis profile offset by $+$30~\kms\ for clarity. Best-fit parameters and their $1\sigma$ uncertainties are listed in the legend of each panel.}
    \label{fig:JAMs}
\end{figure*}

We retrieved the \citet{Fisher2008} composite $V$-band surface brightness profile of NGC~5055, i.e., the central \hst/WFPC2 PC F606W profile mated to the wide-field Sloan Digital Sky Survey (SDSS) profile \citep{York2000}.  We performed an independent decomposition using a Python~3 version of the \textsc{Profiler} code \citep{Ciambur2016}\footnote{The \textsc{Profiler} code was updated to Python3 by A.~W.~Graham.} to fit a three-component parametric model: two Sérsic functions plus an anti-truncated (broken) exponential disk, chosen to provide a good empirical match to the observed profile. The best-fit parameters, shown in  Figure~\ref{fig:ngc5055_f606w_SB}, are: an inner Sérsic component with $\mu_{e,1} = 18.55$~mag~arcsec$^{-2}$, $r_{e,1} = 3\farcs69$, and $n_1 = 5.14$; a middle Sérsic component with $\mu_{e,2} = 20.45$~mag~arcsec$^{-2}$, $r_{e,2} = 30\farcs9$, and $n_2 = 0.8$ (held fixed); and an anti-truncated exponential disk with central surface brightness $\mu_0 = 19.83$~mag~arcsec$^{-2}$, break radius $r_b = 258\farcs9$, and inner and outer scale lengths $h_1 = 83\farcs5$ and $h_2 = 123\farcs8$, respectively. Since all fitted components have effective radii well exceeding the WFPC2/PC PSF FWHM ($\sim$$0\farcs1$), PSF-broadening has a negligible effect on these reported parameters. The overall fit quality is characterized by a root-mean-square residual of $\Delta_{\rm rms} = 0.074$~mag~arcsec$^{-2}$, confirming that the adopted parametric model provides an adequate empirical description of the observed light profile. This decomposition improves upon \citet{Fisher2008} in two respects: it explicitly includes the compact inner component they excluded from their fit, and it adopts an anti-truncated disk model to capture the observed upturn in the profile beyond $r \approx 260\arcsec$.

We stress that these Sérsic components serve purely as a flexible empirical parameterization of the light profile to enable MGE construction; they do not carry a physical interpretation as distinct structural components, for which we refer to Section~\ref{sec:M-galaxy_relation}. The morphological classification of NGC~5055's central region remains debated in the literature; \citet{Fisher2008} consider it to have a pseudobulge with $n < 2$ (dominating from $\sim$4$\arcsec$ to $\sim$70$\arcsec$), while \citet{Afanasiev_Silchenko_2002} favor a classical bulge interpretation (within $\sim$4$\arcsec$)  based on the near-$R^{1/4}$ inner profile. Although, both studies acknowledge complicating factors including a circumnuclear stellar disk, dust lanes, and nuclear spiral structure. Given that \citet{Afanasiev_Silchenko_2002} note that the excess light over the inner few arcseconds stems from, among other things, a circumnuclear stellar disk (confirmed by Figure~\ref{fig:ngc5055_f606w_SB}), we do not regard this excess to be a traditional bulge.

We adopt our three-component parametric fit to the \citet{Fisher2008} composite $V$-band light profile as our fiducial stellar mass model, with the \hst/WFPC2 F547M two-component fit of \citet{Pechetti2020} serving as an independent check to bracket the systematic uncertainty from the choice of photometric data. In both cases, the best-fit parameters were converted into \mge\ models using the \texttt{mge\_fit\_1d} routine \citep{Cappellari2002}, applied to the circularized 1D major-axis profile on a logarithmic radial grid, with the observed axial ratio $q'_j$ of each Gaussian component assigned from the ellipticity measurements of \citet{Pechetti2020}. We fit the MGE to the parametric model rather than the data directly, as the composite profile (HST + ground-based) becomes increasingly noisy at large radii and a smooth analytical representation avoids spurious Gaussian components in the MGE. Both \mge\ models comprise 17 Gaussian components and are listed in Table~\ref{table:mge-table}.

\begin{figure*}[!th]
    \centering
    \includegraphics[width=0.99\textwidth]{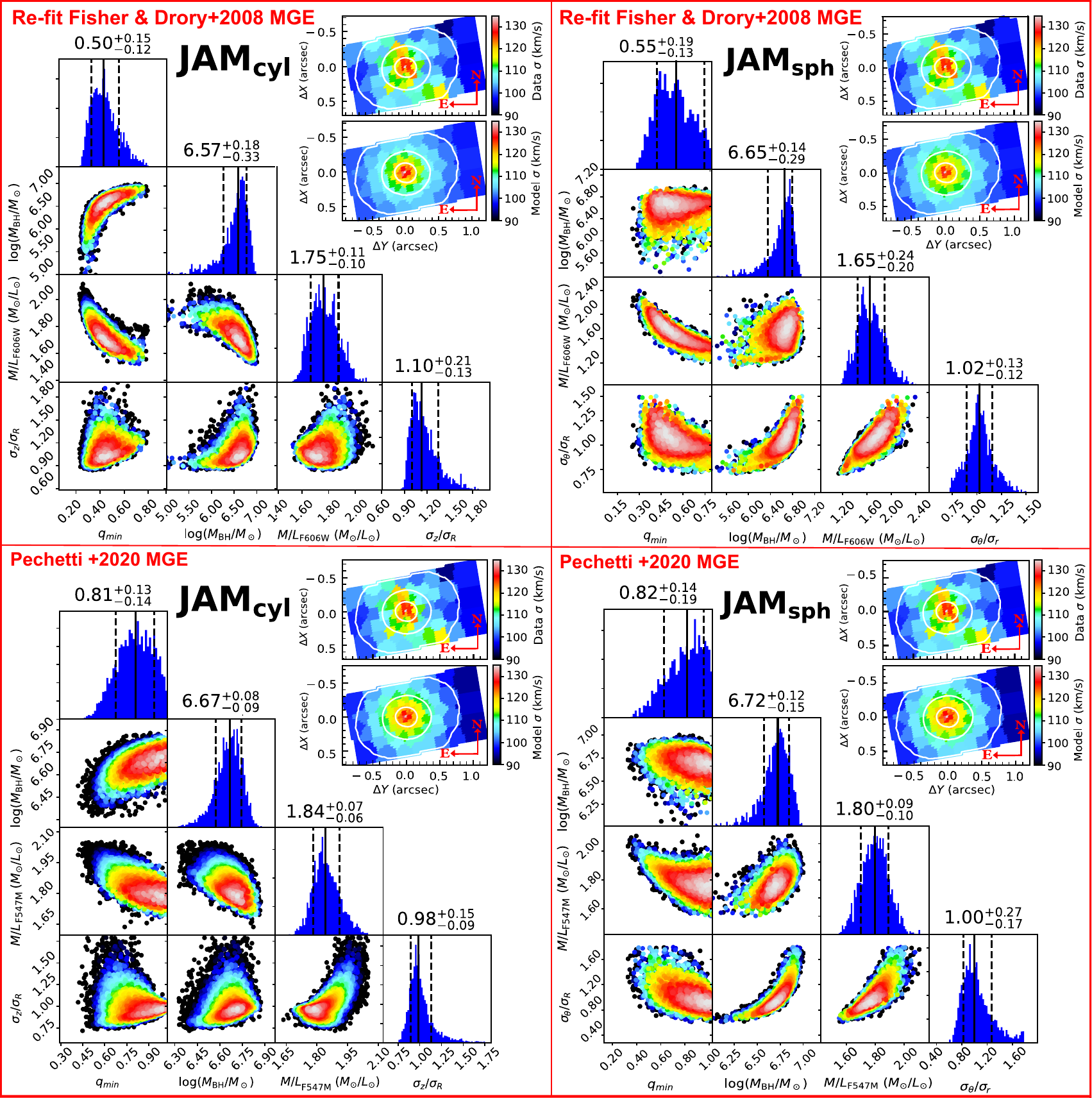}  
    \caption{Posterior distributions from the \adamet\ MCMC optimization of the \jam\ models for the \osiris\ nuclear stellar kinematics of NGC~5055, shown after burn-in.  The corner plot displays the 2D joint and 1D marginalized posterior distributions of \Mbh, \mlhst, $\sigma_z/\sigma_R$, and $i$. Solid and dashed vertical lines mark the best-fit values and $1\sigma$ uncertainties, respectively. \textit{Inset panels:} observed (top) and best-fit model (bottom) $\sigma$ maps on a consistent velocity scale, demonstrating the high fidelity of the fit across the \osiris\ FOV. \textit{Top-row plots:} Results for \jamcyl\ ({\it left}) and \jamsph\ ({\it right}) with the \citet{Pechetti2020} MGE photometric (mass) model.  \textit{Bottom-row plots:} Same as the top-row plots, but with the \citet{Fisher2008} MGE photometric (mass) model.}
    \label{fig:adamet}
\end{figure*}

\begin{figure}
\centering
\includegraphics[width=0.45\textwidth]{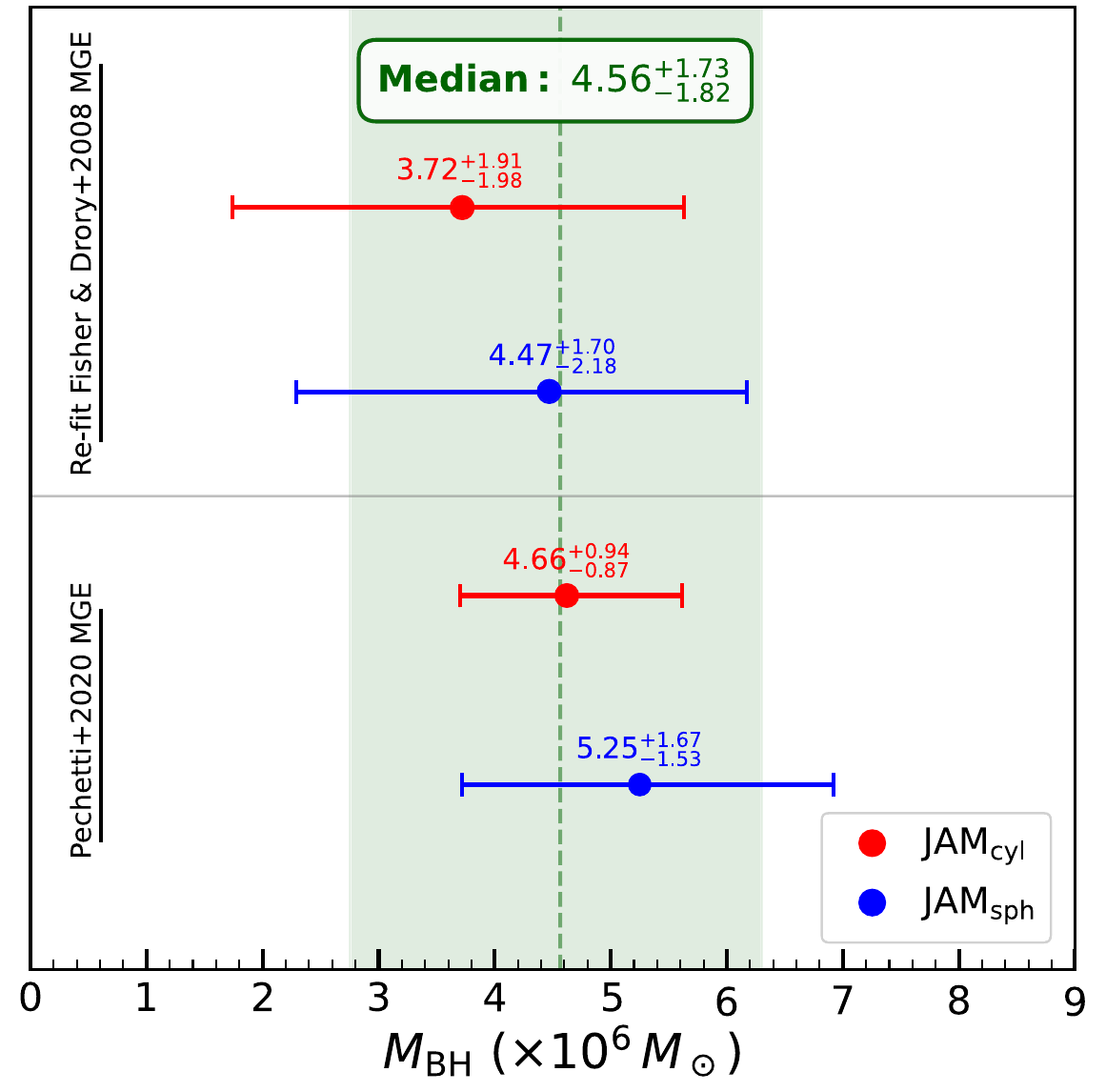}
\caption{Summary of \Mbh\ measurements for NGC~5055 from four \jam\ models applied to the Keck/\osiris\ stellar kinematics, spanning two velocity-ellipsoid alignments (\jamcyl, \jamsph) and two photometric models (F547M, F606W). Individual measurements with their $1\sigma$ uncertainties are shown as points with error bars. The vertical dashed line and shaded region indicate the adopted \Mbh\ and its total quadrature (statistical and systematic) uncertainty, respectively.}
\label{fig:MBHsummary}
\end{figure} 

We note that the MGE parameterization of a surface brightness profile is mathematically non-unique: multiple combinations of Gaussian components can equally well reproduce the observed light distribution to within the measurement uncertainties. The two MGE models presented in Table~\ref{table:mge-table} differ in their individual Gaussian parameters, reflecting this degeneracy rather than any physical distinction between the underlying stellar mass distributions. What matters for the dynamical modeling is that both models accurately reproduce the observed surface brightness profile; as demonstrated by the consistent \Mbh\ values recovered across all four JAM models (Table~\ref{tab:adamet}), the dynamical results are robust against the specific choice of MGE decomposition.

\section{Dynamical Modelling}\label{sec:dynamical-models}

\subsection{Jeans Anisotropic Models}\label{sec:jam}

We modeled the observed $\sigma$ (and $V_{\rm rms}$) maps using both the cylindrically aligned \citep[\jamcyl;][]{Cappellari08} and spherically aligned \citep[\jamsph;][]{Cappellari2020} implementations of JAM, each assuming an axisymmetric potential, to assess how the assumed velocity-ellipsoid alignment affects the dynamical \Mbh\ determination in NGC~5055.

In \jamcyl, the velocity ellipsoid is aligned with cylindrical coordinates $(R,\,\phi,\,z)$, with anisotropy parameterized as $\beta_z \equiv 1 - (\sigma_z/\sigma_R)^2$,  where $\sigma_z \neq \sigma_R = \sigma_\phi$. In \jamsph, the ellipsoid is instead aligned with spherical coordinates $(r,\,\theta,\,\phi)$, with anisotropy parameterized as $\beta_r \equiv 1 - (\sigma_\theta/\sigma_r)^2$, where $\sigma_r \neq \sigma_\theta = \sigma_\phi$. In both cases, all three velocity-dispersion components are permitted to differ. Models were computed using the \texttt{jam\_axi\_proj} routine in \textsc{JamPy} \citep{Cappellari2020}, with \texttt{align=`cyl'} and \texttt{align=`sph'} for the respective alignments, yielding the projected second-moment map $\langle V_{\rm los}^2 \rangle^{1/2} \equiv V_{\rm rms}=$~\Vrms, in which, the intrinsically projected second moments of $\langle\sigma_\theta^2\rangle$ and $\langle\sigma_r^2\rangle$ for \jamsph\ and of $\langle\sigma_z^2\rangle$ and $\langle\sigma_\phi^2\rangle$ for \jamcyl. This approximation is purely a numerical convenience that enables fast model evaluation predictions for \jamcyl, whereas \jamsph\ always computes the BH gravitational potential exactly. 

Both \jamcyl\ and \jamsph\ fit for \Mbh\ (sampled logarithmically) and, linearly, the stellar mass-to-light ratio ($M/L_{\rm F606W}$ or \mlhst), the inclination ($i$, internally converted to the minimum intrinsic axial ratio $q_{\rm min}$), and the orbital anisotropy ($\beta_z$ or $\beta_r$), which was converted into the equivalent dispersion ratios $\sigma_\theta/\sigma_r$ and $\sigma_z/\sigma_R$ for display purposes, respectively. The initial guesses and uniform priors over the search ranges for these four parameters are listed in Table~\ref{tab:adamet}. Given the limited radial extent of the OSIRIS FOV ($\sim$$1'' \times 2''$, or $\sim$$43 \times 86$~pc at $D = 8.87$~Mpc) and the absence of a reliable color map --- the only available HST bands covering the nucleus are F606W and F547M, separated by only $\sim$600~\AA, too small a baseline for a robust constraint on spatial $M/L(r)$ variations --- we thus adopt spatially constant anisotropy and $M/L$ in both models. This assumption is further supported by the DESI nuclear spectrum, which is well fit by a single old ($t \approx 2.3$~Gyr), metal-rich (${\rm [Fe/H]} \approx +0.70$) stellar population with no evidence for multiple components within the $1\farcs5$ fiber (Section~\ref{sec:desi}). The same FOV also precludes a meaningful constraint on radially varying anisotropy, but see Section~\ref{sec:add_uncer} for the impact of testing varying anisotropy profiles ($\beta_{z\,{\rm or}\,r}(r)$) on the \Mbh\ determination.

\begin{table*}
\caption{\jam\ models best-fitting parameters and formal uncertainties} 
\vspace{-4mm}\centering
\begin{tabular}{lcccc|ccc} 
\hline\hline       
Parameter (Unit) & Search range & Best-fit & $1\sigma$ Error& $3\sigma$ Error& Best-fit & $1\sigma$ Error& $3\sigma$ Error\\
(MGE)&  & value & (16--84)\% & (0.14--99.86)\% & value   & (16--84)\% & (0.14--99.86)\% \\  
\multicolumn{1}{c}{(1)} & (2) & (3) & (4) & (5) & (3) & (4) & (5) \\  
\hline 
Refit \citet{Fisher2008}  & F606W & \multicolumn{3}{c}{\jamcyl} & \multicolumn{3}{c}{\jamsph} \\ 
\hline 
$\log$(\Mbh/\Msun)&  $(0\to8)$ &  $6.57$  &$-0.33$, $+0.18$ &$-1.52$, $+0.37$ & $6.65$   & $-0.30$, $+0.14$ & $-1.33$, $+0.30$\\
$M/L_{\rm F606W}$ (\Msun/\Lsun)      &  $(0.1\to3)$  &  $1.75$  &$-0.10$, $+0.11$ &$-0.22$, $+0.31$ & $1.65$   &$-0.20$, $+0.24$  &$-0.45$, $+0.70$\\  
$q_{\rm min}$             &$(0.05\to0.99)$  & $0.50$ &$-0.12$, $+0.15$ &$-0.22$, $+0.42$ & $0.55$ & $-0.13$, $+0.19$ & $-0.27$, $+0.42$\\
($\sigma_z/\sigma_R$)$_{\rm cyl}$ or ($\sigma_\theta/\sigma_r$)$_{\rm sph}$& $(0.5\to1.5)$   & $1.10$  &$-0.13$, $+0.21$ &$-0.24$, $+0.75$ & $1.02$  & $-0.13$, $+0.13$ & $-0.29$, $+0.46$\\[1mm]
\hline
\citet{Pechetti2020}  & F547M & \multicolumn{3}{c}{\jamcyl} & \multicolumn{3}{c}{\jamsph} \\ 
\hline 
$\log$(\Mbh/\Msun)&  $(0\to8)$ &  $6.67$  &$-0.09$, $+0.08$ &$-0.32$, $+0.18$ & $6.72$   & $-0.15$, $+0.12$ & $-0.68$, $+0.26$\\
$M/L_{\rm F547M}$ (\Msun/\Lsun)      &  $(0.1\to3)$  &  $1.84$  &$-0.06$, $+0.07$ &$-0.15$, $+0.20$ & $1.80$   &$-0.10$, $+0.09$  &$-0.24$, $+0.30$\\  
$q_{\rm min}$             &$(0.05\to0.99)$  & $0.81$ &$-0.14$, $+0.13$ &$-0.34$, $+0.21$ & $0.82$ & $-0.19$, $+0.14$ & $-0.44$, $+0.21$\\
($\sigma_z/\sigma_R$)$_{\rm cyl}$ or ($\sigma_\theta/\sigma_r$)$_{\rm sph}$& $(0.5\to1.5)$   & $0.98$  &$-0.09$, $+0.15$ &$-0.23$, $+0.75$ & $1.00$  & $-0.18$, $+0.27$ & $-0.38$, $+0.78$\\[1mm]
\hline
\end{tabular}
\label{tab:adamet}
\end{table*}

In total, we perform four \jam\ model runs spanning two choices of velocity-ellipsoid alignment: cylindrical (\jamcyl) or spherical (\jamsph), and two stellar mass models: our three-component F606W parametric fit and the \citet{Pechetti2020} F547M model (Table~\ref{table:mge-table}), to bracket the dominant sources of systematic uncertainty in our dynamical \Mbh\ determination.

\subsection{MCMC Fitting and Uncertainty Estimation}\label{sec:mcmc_uncertainty}

We sampled the \jam\ parameter space using a Markov chain Monte Carlo (MCMC) approach, employing the adaptive Metropolis algorithm \citep{Haario2001} as implemented in the \adamet\ package\footnote{v2.0.9: \url{https://pypi.org/project/adamet/}} \citep{Cappellari2013a}. Each chain comprised $5\times10^4$ steps, with the first 20\% discarded as burn-in. Best-fit parameters correspond to the maximum-likelihood solution, and uncertainties are quoted as $1\sigma$ (16th--84th percentile) and $3\sigma$ (0.14th--99.86th percentile) confidence levels (CLs) from the posterior distributions.

The large number of high-precision \osiris\ kinematic bins introduces two challenges for the \jam\ modeling. First, formal statistical uncertainties on the fitted parameters can be unrealistically small when unmodeled systematics are present. Second, the $\chi^2$ statistic can be disproportionately driven by the many data points at large radii, potentially biasing \Mbh\ away from the kinematics within the SOI where the BH gravitational potential dominates. 

To account for potential systematic errors in the \Mbh\ measurements due to those two issues, we adopt the $\Delta\chi^2$-inflation scheme of \citet{vandenBosch2009}, which in a Bayesian context is equivalent to scaling the kinematic uncertainties by $(2N)^{1/4}$, where $N$ is the number of kinematic bins \citep[Section~6.1]{Mitzkus2017}. Rather than applying this inflation uniformly, we apply it selectively: formal uncertainties are preserved for all $N_{r \leq R_{\rm SOI}}=3$ bins inside the SOI ($r \leq R_{\rm SOI} \approx 0\farcs 048$), where the gravitational influence of the SMBH dominates the kinematics, while the uncertainties of the $N_{r > R_{\rm SOI}} = 79$ bins outside this radius are inflated by $(2N_{r > R_{\rm SOI}})^{1/4}$. This strategy ensures that \Mbh\ is primarily constrained by the central kinematics while still propagating realistic systematic uncertainties at larger radii. This approach is now standard practice in both molecular gas- \citep[e.g.,][]{Smith2019, Smith2021, North2019, Ngo2025_7052, Nguyen20, Nguyen21, Nguyen22, Nguyen_2026_4061, Nguyen_2026_315} and stellar-based dynamical \Mbh\ measurements \citep[e.g.,][]{Drehmer2015, vandenBosch2009, Thater22, Nguyen_2026_4258, Nguyen_2026_M81}.

To test whether $M_{\rm BH}$ depends on the choice of inflation radius, we ran three additional JAM fits: inflating at $r \geq 0\farcs15$, at $r \geq 0\farcs25$, and excluding all bins at $r > 0\farcs65$ with no inflation applied (69 of 82 bins retained). In each case, the best-fit $M_{\rm BH}$ from all four JAM configurations (JAM$_{\rm cyl}$/JAM$_{\rm sph}$ $\times$ two MGE models) is consistent with the original within the uncertainties, and the median $M_{\rm BH}$ across the four models shifts by less than $1\%$: from $4.56\times10^6\,M_\odot$ (original) to $4.59$, $4.58$, and $4.60\times10^6\,M_\odot$, respectively. The BH mass is insensitive to the specific inflation radius adopted.

\section{Results and Discussion}\label{sec:results_discussion}

\subsection{Supermassive Black Hole Mass}\label{sec:smbh}

The best-fit parameters from all four \jam\ models are summarized in Table~\ref{tab:adamet}. The two \jamcyl\ models yield \Mbh~$= (4.66^{+0.94}_{-0.87})\times10^6$~\Msun\ and $(3.72^{+1.91}_{-1.98})\times10^6$~\Msun\ for the \citet{Pechetti2020} and our refit \citet{Fisher2008} MGEs, respectively, while the two \jamsph\ models yield \Mbh~$= (5.25^{+1.67}_{-1.53})\times10^6$~\Msun\ and $(4.47^{+1.70}_{-2.18})\times10^6$~\Msun. All four measurements are consistent within their $1\sigma$ uncertainties, confirming that \Mbh\ is insensitive to both the assumed velocity-ellipsoid alignment and the choice of photometric model. However, the \jamsph\ models systematically return slightly larger \Mbh\ values and broader uncertainties than their \jamcyl\ counterparts, consistent with the greater degeneracy between the radial anisotropy parameter $\beta_r$ and \Mbh\ in spherically aligned models \citep[e.g.,][]{Cappellari2020, Thater22, Nguyen_2026_M81}.

The sphere of influence, $R_{\rm SOI} = GM_{\rm BH}/\sigma_e^2 \approx 0\farcs048$, is only marginally resolved with respect to the PSF FWHM of $0\farcs045$ (Figure~\ref{fig:star_psf_mge}), falling below the factor-of-two criterion commonly adopted as a threshold for reliable \Mbh\ measurements \citep[e.g.,][]{Ferrarese2005}. We note, however, that the JAM models are fitted simultaneously to all 82 spatial bins across the OSIRIS FOV; the constraint on $M_{\rm BH}$ arises from the overall shape of the 2D stellar velocity dispersion profile; specifically, the central $\sigma$ rise that models with significantly lower $M_{\rm BH}$ fail to reproduce, rather than from the spatial resolution of the SOI alone. The robustness of this constraint is further demonstrated by seven independent systematic tests in Section~\ref{sec:add_uncer}, all of which yield $M_{\rm BH}$ values well within $1\sigma$ of the fiducial result.

The larger \Mbh\ uncertainties from our refit \citet{Fisher2008} MGE relative to \citet{Pechetti2020} are primarily driven by its lower and less well-constrained $q_{\rm min} \approx 0.5$--$0.55$, which introduces greater degeneracy between inclination, $M/L$, and \Mbh. This $q_{\rm min}$ corresponds to $i \approx 67\degr$, in closer agreement with the large-scale \ion{H}{1} disk inclination \citep[$i \approx 59\degr$;][]{Battaglia_2006}, as expected given that the \citeauthor{Fisher2008} profile extends to $\approx$ 660\arcsec and is thus sensitive to the large-scale disk geometry. In contrast, the \citet{Pechetti2020} MGE yields $q_{\rm min} \approx 0.81$--$0.82$, exceeding the observed galaxy axial ratio ($q' = 0.61$), indicating that the JAM model finds an intrinsically round nuclear stellar distribution within the inner $\approx$10\arcsec, consistent with a compact, nearly circular nuclear component embedded within the more flattened outer disk, and precluding a meaningful inclination estimate from this model alone.

Similar to the derived \Mbh, all models also consistently recover nearly isotropic stellar orbits ($\beta_z \approx \beta_r \approx 0$), with $\sigma_z/\sigma_R \approx$~0.98--1.10 for \jamcyl\ and $\sigma_\theta/\sigma_r \approx$~1.00--1.02 for \jamsph. This is consistent with results from \jwst/NIRSpec \citep{Nguyen_2026_4258, Nguyen_2026_M81} and VLT/MUSE \citep{Simon2024} stellar dynamical studies of similar systems, and with the mild-to-near-isotropic anisotropy commonly found in rotating galaxies from \jam\ modeling \citep{Cappellari2007, Cappellari08}.

\begin{figure*}[!th]
    \centering
    \includegraphics[width=0.95\textwidth]{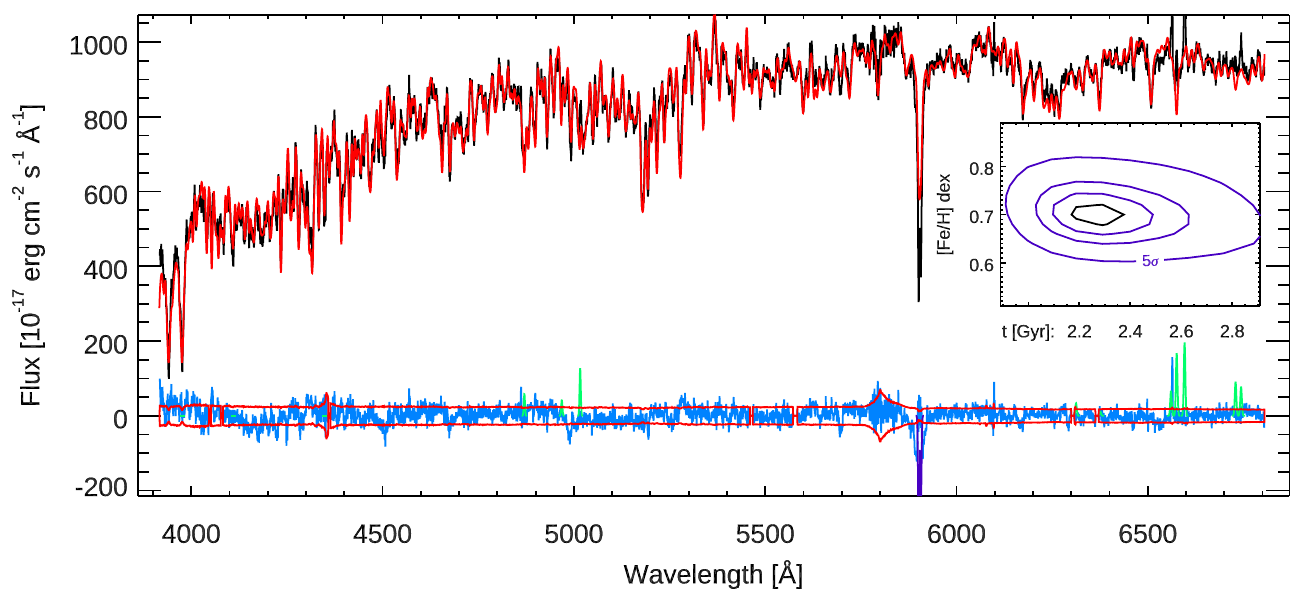}  
    \caption{DESI DR1 nuclear spectrum of NGC~5055 extracted within a $1\farcs5$-diameter fiber aperture, fitted with PEGASE.HR stellar population models \citep{LeBorgne2004} using \ppxf. The observed spectrum (black) is overlaid with the best-fit stellar population model (red); fitting residuals are shown in blue; emission lines in green and the NaD absorption in violet. \textit{Inset:} $\chi^2$ confidence contours in the stellar age--metallicity plane, shown at 1 (black), $2,3,5\sigma$ (blue) levels. The well-constrained minimum at $t \approx 2.3$~Gyr and $\mathrm{[Fe/H]} \approx +0.70$ confirms a moderately old, highly super-solar nuclear stellar population.}
    \label{fig:desi}
\end{figure*}

The dynamical $M/L$ values are consistent between the two velocity-ellipsoid alignments, with \mlhst~$\approx 1.80$--$1.84$~\Msun/\Lsun\ for the \citet{Pechetti2020} MGE and $M/L_{\rm F606W} \approx 1.65$--$1.75$~\Msun/\Lsun\ for our refit \citet{Fisher2008} MGE, the latter being somewhat lower as expected from the redder passband of F606W relative to F547M. The independent stellar population analysis of the DESI nuclear spectrum, yielding $M/L \approx 1.9$~\Msun/\Lsun\ for a $\sim$$2.26\pm0.08$~Gyr, supersolar-metallicity population, is in excellent agreement with our dynamical $M/L$ and provides stronger external validation (see Figure~\ref{fig:desi}).

Figure~\ref{fig:JAMs} shows the \osiris\ velocity dispersion ($\sigma$) profiles of NGC~5055 extracted along the major and minor axes, with error bars inflated as described in Section~\ref{sec:mcmc_uncertainty}, overlaid with the best-fit \jamcyl\ and \jamsph\ models (corresponding to the two adopted MGE stellar mass models for NGC~5055). All four models provide a good match to the observed $\sigma$ profiles along both axes across the \osiris\ FOV.

Figure~\ref{fig:adamet} shows the 2D posterior distributions from the \adamet\ MCMC optimization for all four \jam\ models.  Point colors indicate relative likelihood, with white denoting the maximum-likelihood solution and the $1\sigma$ CL, and black the $3\sigma$ boundary. The 1D marginalized histograms along the diagonal yield the best-fit parameter values and $1\sigma$ uncertainties, incorporating both stellar kinematic and statistical errors. The inset velocity dispersion maps in the upper-right corner of each panel compare the observed and best-fit model, with the map plotted against the world coordinate system (WCS) orientation: north is up and east to the left. All four models reproduce the observed kinematics to within 6\% relative residuals across the \osiris\ FOV.

In the \jamcyl\ models, \Mbh\ and $M/L$ exhibit a \textit{negative} covariance: a higher $M/L$ assigns more mass to the stellar component, reducing the BH contribution needed to reproduce the observed $\sigma$ peak. This classic trade-off is a well-known systematic in stellar dynamical modeling \citep{Cappellari08, Thater22}, and vice versa. In contrast, the \jamsph\ models exhibit a \textit{positive} \Mbh--$M/L$ covariance, a consequence of the spherically aligned velocity ellipsoid: the radial anisotropy naturally promoted by this geometry inflates the projected line-of-sight $\sigma$ such that both \Mbh\ and $M/L$ must increase together to reproduce the observed kinematics, rather than trading off against each other.

A \textit{positive} covariance between the anisotropy parameter ($\sigma_z/\sigma_R$ for \jamcyl\ and $\sigma_\theta/\sigma_r$ for \jamsph) and both \Mbh\ and $M/L$ is consistently observed across all four models, reflecting the classical mass--anisotropy degeneracy \citep{Binney1982, Gerhard1993}. More isotropic orbits produce a weaker central $\sigma$ peak, requiring a higher \Mbh\ to reproduce the observed kinematics; simultaneously, more radially biased orbits inflate the line-of-sight $\sigma$, which can equivalently be compensated by increasing $M/L$, making anisotropy, \Mbh, and $M/L$ difficult to disentangle from kinematics alone. This degeneracy represents an additional source of systematic uncertainty in our \Mbh\ determination, partially mitigated by the fact that all four models consistently prefer nearly isotropic orbits ($\sigma_z/\sigma_R \approx \sigma_\theta/\sigma_r \approx 1$), limiting the range of allowed anisotropy and thus the associated uncertainty on \Mbh. Additionally, $q_{\rm min}$ exhibits a consistent \textit{negative} covariance with $M/L$: a rounder (more face-on) distribution produces weaker line-of-sight projection of the stellar velocity field, reducing the stellar mass contribution needed to reproduce the observed kinematics and thus requiring a lower $M/L$.

Figure~\ref{fig:MBHsummary} shows our four \Mbh\ measurements and their $1\sigma$ uncertainties from the four conventional \jam\ models, providing a summary of the most precise \Mbh\ constraint for NGC~5055 based on the Keck/\osiris\ stellar kinematics. We derive a robust SMBH estimate by adopting the median and its $1\sigma$ statistic-bootstrapping CL across the four models, yielding \Mbh~$= (4.56 ^{+0.39}_{-0.47})\times10^6$~\Msun

To construct a comprehensive uncertainty budget, we combine in quadrature two contributions: (1) the statistical uncertainty from the MCMC posterior of each individual \jam\ model, which incorporates the propagated stellar kinematic measurement errors (including the selective error inflation described in Section~\ref{sec:mcmc_uncertainty}); and (2) the systematic uncertainty estimated from the model-to-model spread across the four \jam\ runs, reflecting variations in the assumed velocity-ellipsoid alignment and stellar mass model. Following common practice in dynamical \Mbh\ determinations, the uncertainty due to the assumed distance is excluded from this budget, as \Mbh\ scales linearly with distance and this contribution is treated separately (Section~\ref{sec:add_uncer}). This yields our final adopted mass of \Mbh~$= (4.56 ^{+1.73}_{-1.82})\times10^6$~\Msun\ for NGC~5055.

Our dynamically measured \Mbh\ is 30\% lower than the value predicted by the \citet{Sahu19b} \Mbh--$\sigma$ relation (Section~\ref{sec:targets}), which is not surprising given that scaling relations carry intrinsic scatter of $\sim$0.3--0.5~dex and are calibrated predominantly on ETGs with classical bulges, while the morphological classification of NGC~5055's central structure remains uncertain (Section~\ref{sec:photomodel}). This stellar-dynamical \Mbh\ yields $R_{\rm SOI} \approx 0\farcs048$ ($\approx$2.06~pc), comparable to the Keck/\osiris\ PSF FWHM, confirming that the gravitational influence of the SMBH is detected within the central resolution element of our data.

\begin{figure*}
\centering
\includegraphics[height=0.25\textwidth]{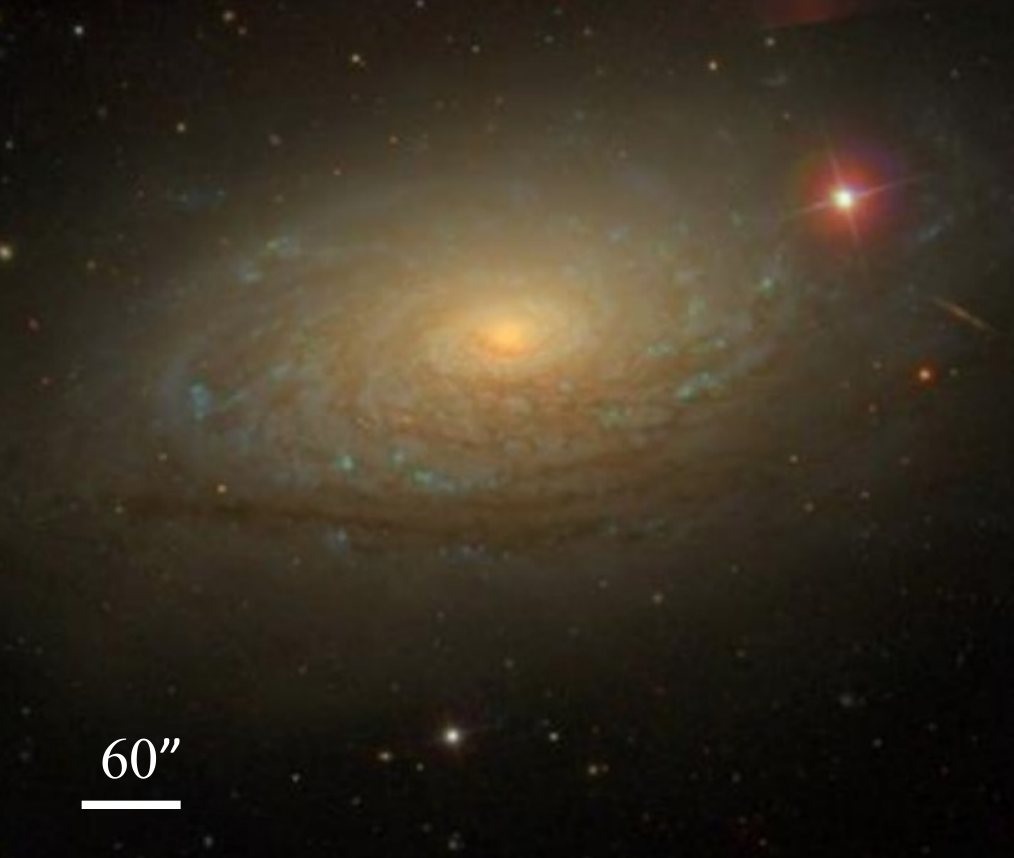}
\includegraphics[height=0.25\textwidth]{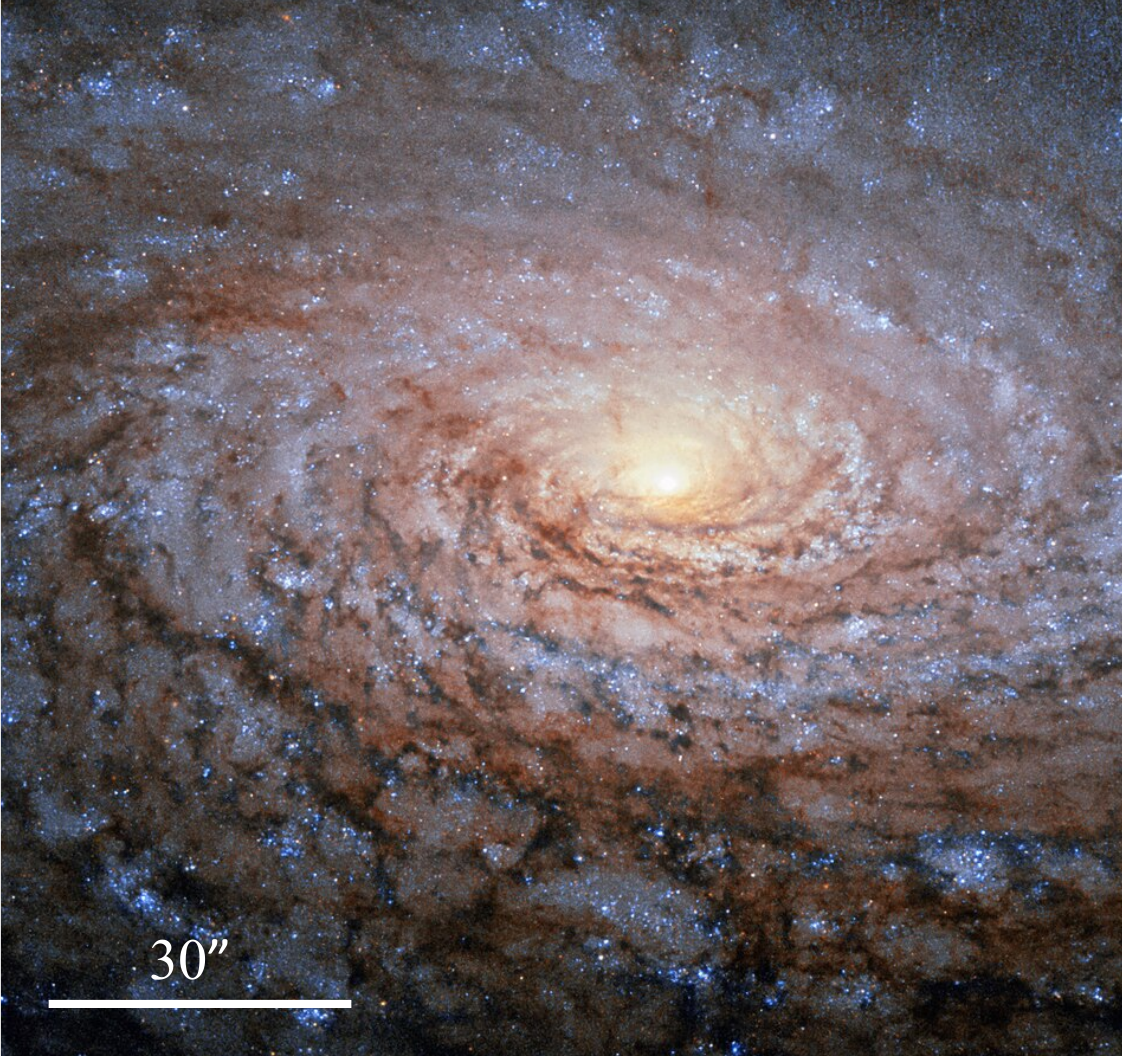}
\includegraphics[height=0.25\textwidth]{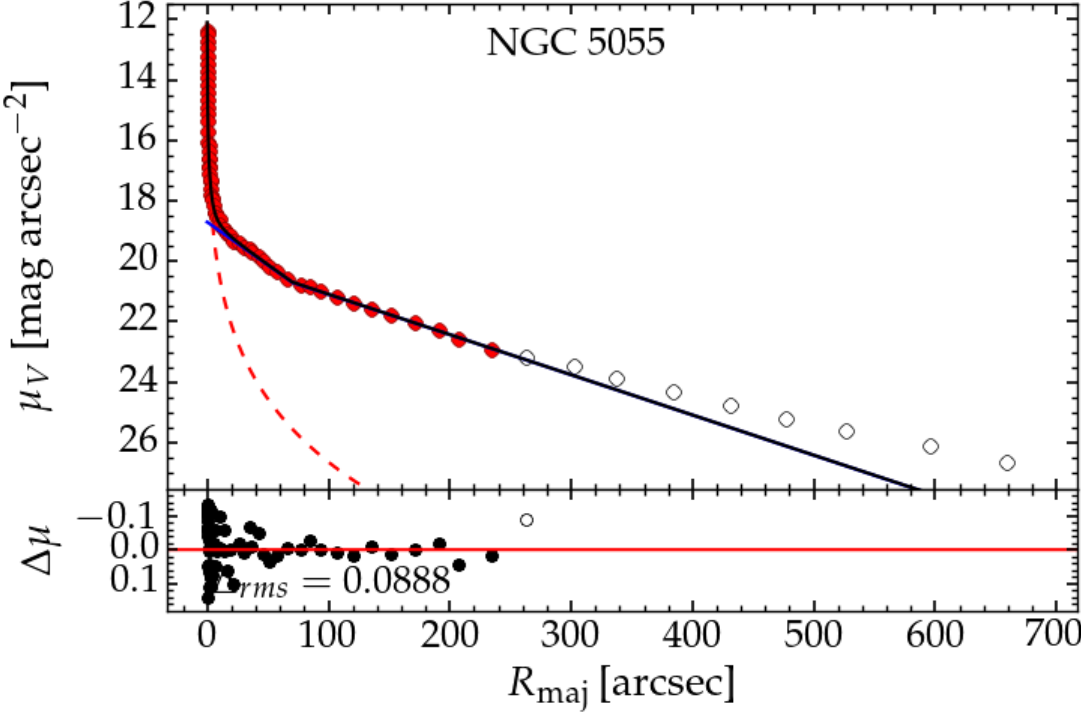}
\caption{{\it Left:} {\it SDSS} $gri$ DR8 image of NGC~5055 \citep{Aihara2011}\footnote{\url{https://skyserver.sdss.org/dr8/en/tools/chart/navi.asp}}. {\it Middle:} \hst\ image (Credit: ESA/Hubble \& NASA).  
{\it Right:} Variant of Figure~\ref{fig:ngc5055_f606w_SB} using a linear radial scale and a two-component fit.  An inner S\'ersic component (dashed red line) again approximates the complexities of the central structure while a revised anti-truncated exponential disk (solid blue line) reveals a disk displaying a double anti-truncated behaviour.}
\label{fig:N5055_V_SB}
\end{figure*}

\subsection{Independent Stellar Population Validation from DESI Spectroscopy}
\label{sec:desi}

We obtained an independent constraint on the stellar \ml$_V$ of the NGC~5055 nucleus from a high S/N spectrum provided by the Dark Energy Spectroscopic Instrument \citep[DESI;][]{DESICollaboration2026} Data Release~1 (DR1), obtained within a circular fiber aperture of diameter~$1\farcs5$ centered on the nucleus. The spectrum was fitted with the PEGASE.HR stellar population models \citep{LeBorgne2004} adopting the Kroupa initial mass function \citep[IMF;][]{Kroupa2001}, using the updated {\sc NBursts} full spectrum fitting \citep{Chilingarian2007a, Chilingarian2007b, Pulatova2025}. The model included a stellar population template, emission lines (H$\alpha-\epsilon$, [O{\sc iii}] 4959/5007~\AA, [O{\sc i}] 6300/6364~\AA, [N{\sc ii}] 6548/6584~\AA, [S{\sc ii}] 6717/6731~\AA) with the flux ratios of forbidden oxygen and nitrogen doublets fixed to the values from the NIST atomic physics database, and additional absorption components in the neutral sodium doublet (NaD). All three components have independent kinematics ($v, \sigma$).

We note that the more commonly used XSL \citep{Verro2022b} and eMILES \citep{Vazdekis2016} stellar population template grids leave substantial spectral residuals around $\sim$7300~\AA, indicative of the spectrum reaching the upper metallicity boundary of those grids ([Fe/H]$=+0.2$~dex for XSL and $+0.4$ for MILES). PEGASE.HR, while lacking coverage redward of $\sim$6800~\AA, yields a significantly better fitting quality over the range 3900-6800~\AA\ and extends to higher metallicities, making it the preferred choice for this nucleus with the super-solar metallicity.

The best-fit stellar population parameters are: a luminosity-weighted age of $2.26 \pm 0.08$~Gyr and a metallicity of $\mathrm{[Fe/H]} = +0.70 \pm 0.05$, yielding a $V$-band stellar mass-to-light ratio of $(M/L)_V = 1.90 \pm 0.1$~\Msun/\Lsun. This is in excellent agreement with our dynamical values of $M/L_\mathrm{F547M} \approx 1.80$--$1.84$ and $M/L_\mathrm{F606W} \approx 1.65$--$1.75$~\Msun/\Lsun\ from the JAM models (Table~\ref{tab:adamet}), providing strong independent validation of the stellar\ml\ recovered from our dynamical modeling.

The DESI spectrum also yields three distinct velocity dispersion measurements from different kinematic tracers: the stellar continuum gives $\sigma_\star = 141 \pm 2$~\kms, the emission lines give $\sigma_\mathrm{em} = 89 \pm 6$~\kms, and the NaD absorption gives $\sigma_\mathrm{NaD} = 92 \pm 3$~\kms. The stellar $\sigma_\star = 141$~\kms\ exceeds the OSIRIS global dispersion of $108 \pm 3$~\kms\ (Section~\ref{sec:optimal_template}), which is trivially explained by the beam-smearing effect: the galaxy rotation within the DESI aperture broadens both absorption and emission lines. The significantly lower dispersions of the emission lines and NaD feature relative to the stellar continuum indicate that these two tracers likely originate in the thin disk component and do not participate in the bulk stellar kinematics. The NaD absorption, in particular, arises primarily from neutral atomic gas along the line of sight \citep[e.g.,][]{Poznanski2012} rather than from the stellar photosphere, and is therefore not a reliable tracer of stellar dynamics in this system. The ratio of equivalent widths of the NaD lines in the doublet (1.78$\pm$0.08 and 1.49$\pm$0.09~\AA) indicates partial saturation of the 5889~\AA\ line because the ratio is less than $2$, the value expected in the non-saturated case. The measured equivalent width of the fainter 5896~\AA\ line translates to the column density $N_H=(4.8\pm0.3)\times10^{20}$~cm$^{-2}$ \citep[see, e.g.,][]{Draine2011} and $A_V=0.22\pm0.02$~mag and explains why we do not observe substantial extinction in the optical images of NGC~5055 in its nuclear region.

\subsection{Additional Uncertainty Budgets}\label{sec:add_uncer}

{\it Distances:} The distance to NGC~5055 is well constrained with $D = 8.87 \pm 0.31$~Mpc from the TRGB measurement of \citet{McQuinn_2017}, which represents one of the most reliable distance indicators for nearby galaxies. Nevertheless, the NASA/IPAC NED lists 27 independent distance determinations spanning 4.4--10.9~Mpc, reflecting the range of methods and their associated systematics. Since \Mbh~$\propto D$, this distance uncertainty propagates directly and linearly into our black hole mass, yielding a systematic range of 2.26--5.60$\times10^6$~\Msun, which represents a dominant component of the total \Mbh\ error budget.

{\it Stellar Kinematic Measurements with PHOENIX Synthetic Library:} As an additional systematic check, we re-extracted the stellar kinematics  of NGC~5055 using the PHOENIX synthetic stellar library \citep{Husser2013} in place of the fiducial XSL library. The two sets of kinematics agree to within 2\% (Section~\ref{sec:kinemap}), and the resulting differences in \Mbh\ and \mlhst\ are negligible.

{\it Fixed $M/L$:} We also ran four JAM models with $M/L$ fixed at $1.9$, the value measured from the DESI spectrum (Section~\ref{sec:desi}). The median $M_{\rm BH}$ across the four models shifts from $4.56\times10^6\,M_\odot$ (free $M/L$) to $4.40\times10^6\,M_\odot$ ($\sim$$3.5\%$), within $1\sigma$ of the original. $M_{\rm BH}$ is insensitive to whether $M/L$ is held fixed at the DESI value or left as a free parameter.

{\it Varying Anisotropy Profile:} We replace the constant anisotropy with a logistic $\beta(r)$ profile \citep{Simon2024, Nguyen_2026_4258, Nguyen_2026_M81} for the \jamsph\ model and a radially varying anisotropy profile \cite[Eq.~37 of][]{Cappellari2026_jam} for the \jamcyl\ model. Across all four JAM models, $M_{\rm BH}$ shifts by $\lesssim$4\%, well within the 1$\sigma$ uncertainties.

\subsection{\texorpdfstring{The \Mbh\--Galaxy Scaling Diagrams}{The MBH--Galaxy Scaling Diagrams}}\label{sec:M-galaxy_relation}

\begin{figure*}
\centering
\includegraphics[width=0.9\textwidth]{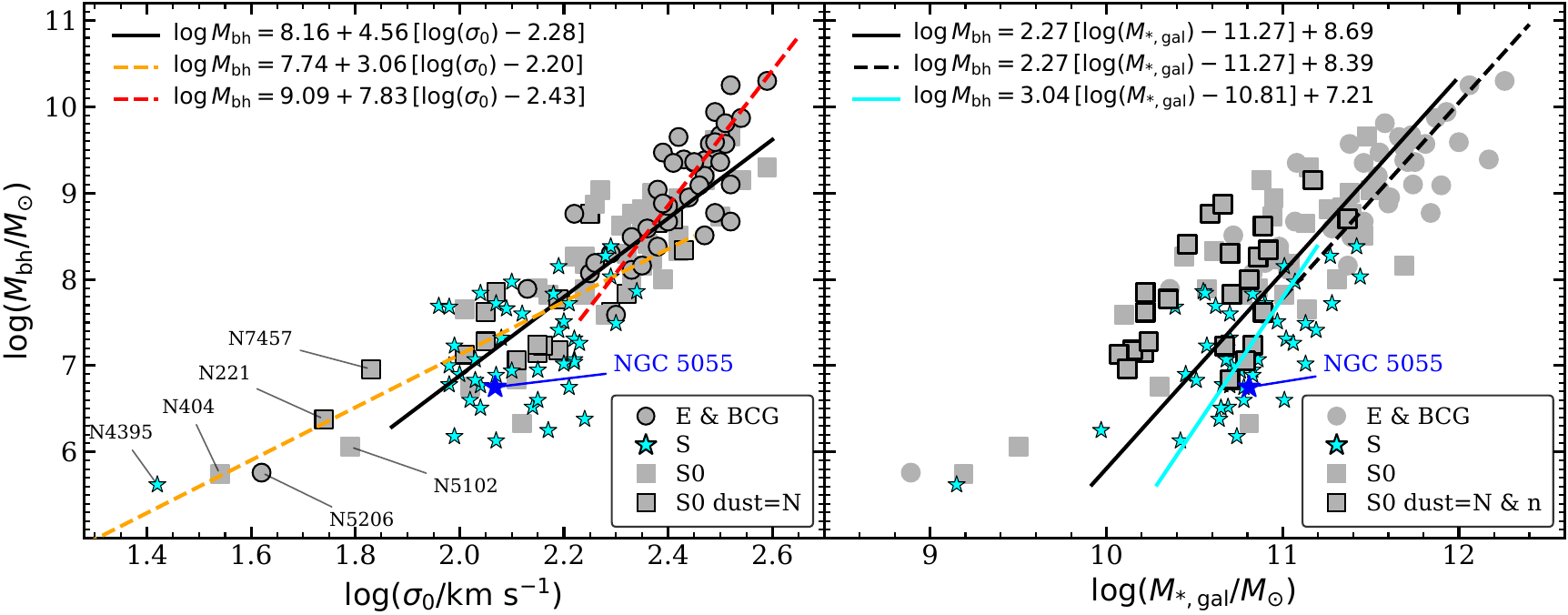}
\caption{The masses are from \citet{Graham2026a} and the velocity dispersions from \citet{Sahu19b}. The equations in the left-hand panel are from \citet{Graham2026b}, and pertain to elliptical (E) galaxies (including brightest cluster galaxies (BCGs): steep dashed line), lenticular (S0) galaxies (solid line), and S0 galaxies with no visible signs of dust (shallow dashed line). 
The named galaxies were excluded from the regression analyses. 
The equations in the right-hand panel are from \citet{Graham2023a} and pertain to galaxies built from major mergers (E, BCG, and dust-rich S0 galaxies: solid line), BCGs alone (parallel dashed line), and spiral (S) galaxies. With the exception of a few suspected faded S galaxies in the lower-right, the dust-poor S0 galaxies (dust $=$ N and n) tend to reside to the low $M_{\rm \star,gal}/M_{\rm BH}$, and thus high $M_{\rm BH}/M_{\rm \star,gal}$, side of the relation defined by the local galaxies containing SMBHs, as noted in \citet{Graham2025}.
NGC~5055 falls $\sim$0.7~dex below the relations defined by elliptical galaxies and AGN hosts, reminiscent of the Milky Way's SMBH which has similarly been noted as undermassive relative to these relations \citep{Zou2024, Lian2025}.}
\label{fig:scaling}
\end{figure*}

\begin{figure}
\centering
\includegraphics[width=1.0\columnwidth]{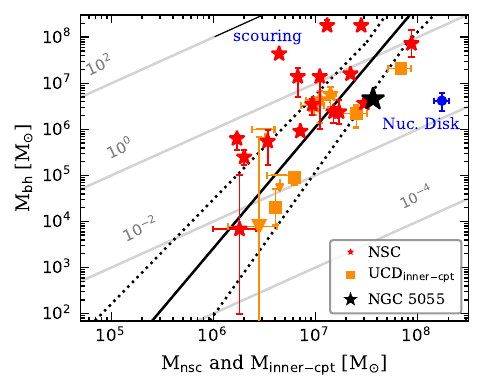}
\caption{\Mbh--\Mnsc\ scaling diagram adapted from \citet{Graham2020} with updates from \citet{Graham2025}. Massive merger-built elliptical (and ellicular) galaxies with $M_{\rm BH}>10^8$~M$_\odot$ are excluded given that their NSCs, if they still possess one, are expected to have been eroded by the merging process.  Flattened nuclear disks can be many tens of parsecs in size \citep[e.g.,][]{Balcells2007}, and are not expected to follow the relation shown here. The two systems with the highest black mass belong to NGC~4026 and NGC~4697.}
\label{fig:nsc-bh}
\end{figure}

While NGC~5055 contains a prominent excess over the inner few arcseconds, brightening the surface brightness by more than $\sim$6~mag~arcsec$^{-2}$, we regard the galaxy as bulgeless. 
%
%
As shown in Figure~\ref{fig:N5055_V_SB}, traversing the light profile from large to small radii, the disk brightens at an accelerating pace (anti-truncates) upon approaching the outer domain ($\sim$260\arcsec) of the weak grand design spiral (Section~\ref{sec:targets}), and anti-truncates again at $\sim$70$\arcsec$ upon approaching the tightly-wound ring-like spiral arms visible in the middle panel of Figure~\ref{fig:N5055_V_SB} \citep[see also][]{Thornley1997}. The central excess is therefore attributed to the superposition of these nested disk substructures, a circumnuclear stellar disk, nuclear spiral arms, and the inner anti-truncated disk, rather than a classical or pseudobulge component. NGC~5055 thus joins a small sample of bulgeless spirals with direct dynamical \Mbh\ determinations, including NGC~2748 \citep{Atkinson2005} and NGC~4395 \citep{denBrok15}, extending the BH mass census into a regime where scaling relations remain poorly constrained and each new measurement serves as a valuable anchor.

Using a (Galactic extinction)-corrected \citep{Schlafly11} $B-V$ colour of 0.704 \citep{deVaucouleurs1991} obtained via the NASA Extragalactic Database (NED)\footnote{\url{https://ned.ipac.caltech.edu/}}, gives a 3.6~$\mu$m diet-Salpeter \citep{Bell01} mass-to-light ratio of 0.46 \citep{Graham2023, Graham2024}.  Applying this to the 3.6~$\mu$m galaxy apparent magnitude of 7.89 AB-mag \citep{Davis2018}, and using a distance modulus of 29.74 mag, and an absolute 3.6~$\mu$m magnitude for the Sun of 6.02 mag (AB), gives a stellar mass $\log(M_{\rm \star,gal}/$\Msun) = 10.81~dex.  This agrees with the value of 10.78~dex reported by \citet{Licquia2015} and \citet{Morales2018}, and 10.81$\pm$0.10~dex reported by \citet{Davis2018}.

Given that the intermediate-scale S\'ersic component that was fit in Figure~\ref{fig:ngc5055_f606w_SB} does not describe a bulge, but rather the inner part of the disk, we show NGC~5055 in the $M_{\rm bh}$--$M_{\rm \star,gal}$ diagram (Figure~\ref{fig:scaling}), but not the $M_{\rm bh}$--$M_{\rm \star,bulge}$ diagram. NGC~5055 can be seen to reside within the cloud of spiral galaxies in both this diagram and the $M_{\rm bh}$--$\sigma_0$ diagram (Figure~\ref{fig:scaling}), revealing nothing unusual with the black hole mass derived herein. Notably, NGC~5055 lies $\sim$0.7~dex below the \Mbh\ scaling relations defined by elliptical galaxies and AGN hosts, placing it in a similar position to the Milky Way, whose central SMBH has also been noted as undermassive relative to these relations \citep{Zou2024, Lian2025}. This offset likely reflects a morphological dependence of the \Mbh--$M_{\star, \rm gal}$ relation, with late-type galaxies and bulgeless systems systematically falling below the locus defined by early-type hosts.

Despite lacking a stellar bar and the associated (peanut shell)-shaped structure of a buckled bar, the Sbc \citep{deVaucouleurs1991} galaxy NGC~5055 is remarkably similar in mass to the Milky Way (an SBbc galaxy), which has $\log (M_{\rm \star,gal}/{\rm M}_\odot)=10.81$~dex \citep{Lian2025} and $M_{\rm BH}=4.0\times10^6$~\Msun\ \citep[][cf., $4.56\times10^6$~\Msun\ for NGC~5055]{EHT_MW_BH}.  Furthermore, the mass of the nuclear star cluster in the Milky Way is $3.0\times10^7$~\Msun\ \citep[][]{Launhardt2002}, which also compares closely with that in NGC~5055 \citep[$3.66\times10^7$~\Msun;][]{Pechetti2020}. As shown in Figure~\ref{fig:nsc-bh}, NGC~5055, intriguingly, serves as yet another system following the \Mbh--\Mnsc\ scaling relation \citep{Graham2016}, prior to the anticipated star cluster erosion in dry galaxy mergers \citep{Bekki2010}. The robustness of this relation is of critical importance for forecasting the rates of extreme-mass-ratio inspiral (EMRI) events, wherein compact stellar remnants (e.g., neutron stars and stellar-mass black holes) residing within the dense nuclear star cluster emit gravitational waves  as they are captured by the central massive black hole \citep[e.g.,][]{Amaro-Seoane2017}.

\section{Conclusions}\label{sec:conclusions}

We have presented the first stellar-dynamical measurement of the SMBH mass in the nearby spiral galaxy NGC~5055 (M63) using LGS-AO IFU with Keck/\osiris. The unprecedented high-angular-resolution (PSF FWHM of 0\farcs045, corresponding to 1.9~pc at $D = 8.87$~Mpc) NIR observations resolve the 2D stellar kinematics that are robust against dust obscuration in the nucleus. Our main results are as follows:

\begin{enumerate}

    \item \textbf{High-resolution stellar nuclear kinematics.} The 2D LOSVDs extracted from the CO bandheads ($\sim$2.3~\micron) via \ppxf\ reveal a velocity dispersion rising from $\sim$90~\kms\ at large radii to a central peak of $\sim$135~\kms\ within $r \lesssim 0\farcs1$, providing unambiguous kinematic evidence for a central dark mass.

    \item \textbf{Robust black hole mass determination.} We modeled the observed kinematics with an ensemble of four \jam\ models spanning two velocity-ellipsoid alignments (\jamcyl\ and \jamsph) and two independent photometric decompositions. All variants yield consistent \Mbh\ values within $1\sigma$, and the median across models gives a fiducial black hole mass of \Mbh~$= (4.56^{+1.73}_{-1.82})\times10^6$~\Msun, with the SOI ($R_{\rm SOI} \approx 0\farcs048$, or $\approx$2.06~pc) resolved by the Keck/\osiris\ PSF.

    \item \textbf{Stellar orbital structure.} Both \jamcyl\ and \jamsph\ consistently recover nearly isotropic stellar orbits in the nuclear region, with $\sigma_z/\sigma_R \approx 0.98$--$1.10$ ($\beta_z \approx 0$) for \jamcyl\ and $\sigma_\theta/\sigma_r \approx 1.00$--$1.02$ ($\beta_r \approx 0$) for \jamsph, consistent with the near-isotropic orbital structure found in stellar dynamical studies of rotating LTGs \citep[e.g.,][]{Cappellari08, Nguyen_2026_M81}.
    
    \item \textbf{Photometric robustness.} Stellar mass models from two independent \hst\ photometric decompositions \citep[][our refit]{Pechetti2020, Fisher2008} yield \Mbh\ values consistent within 14\%, confirming the robustness of our measurement against the choice of photometric model.

    \item \textbf{Image decomposition.} Improvements of either two- or three-component parametric fit to the galaxy's composite $V$-band surface brightness profile disfavors the presence of a classical bulge component (Figures~\ref{fig:ngc5055_f606w_SB} and \ref{fig:N5055_V_SB}, respectively), suggesting that NGC~5055 may be bulgeless, a morphological classification that, if confirmed, would place it among the rare spirals with a direct dynamical \Mbh\ measurement in the absence of a classical bulge.
    
   \item \textbf{Black hole mass scaling relations.} Our black hole mass measurement for NGC~5055 places it squarely within the cloud of data points for other spiral galaxies in the \Mbh--$M_{\rm \star,gal}$ and \Mbh--$\sigma$ diagrams (Figure~\ref{fig:scaling}). It also follows the \Mbh--\Mnsc\ scaling relation, similar to our Milky Way, reinforcing the secular-evolution between the BH and the NSC in NGC~5055.

\end{enumerate}

\section*{Acknowledgements}
DDN acknowledges support from the ELT postdoctoral research fellowship at the Department of Astronomy, University of Michigan. RS acknowledges support from the INAF grant number 1.05.23.04.04. IC’s research is supported by the SAO Telescope Data Center. He also acknowledges support from the NASA ADAP-22-0102 and HST-GO-16739 grants. 

Some of the data presented herein were obtained at Keck Observatory, which is a private 501(c)3 non-profit organization operated as a scientific partnership among the California Institute of Technology, the University of California, and the National Aeronautics and Space Administration. The Observatory was made possible by the generous financial support of the W. M. Keck Foundation. 
\facility{Keck/\osiris.} Keck Program ID: 2024B\_W294.
The authors wish to recognize and acknowledge the very significant cultural role and reverence that the summit of Maunakea has always had within the Native Hawaiian community. We are most fortunate to have the opportunity to conduct observations from this mountain.

\software{
{\tt Python~3.12:} \citep{VanRossum2009}, 
{\tt Matplotlib~3.6:} \citep{Hunter2007}, 
{\tt NumPy~1.22:} \citep{Harris2020}, 
{\tt SciPy~1.3:} \citep{Virtanen2020},  
{\tt photutils~0.7:} \citep{bradley2024}, 
{\tt AstroPy~5.1} \citep{AstropyCollaboration2022}, 
{\tt AdaMet 2.0} \citep{Cappellari2013a}, 
{\tt JamPy~7.2} \citep{Cappellari2020}, 
{\tt MgeFit~5.0} \citep{Cappellari2002}, and 
{\tt Profiler} \citep{Ciambur2016}.
}

\bibliographystyle{aasjournalv7}
\bibliography{NGC5055}

\label{lastpage}
\end{document}